\documentclass[mca,article,accept,a4paper,moreauthors,pdflatex]{Definitions/mdpi_arXiv} 

\firstpage{1} 
\pubvolume{xx}
\issuenum{1}
\articlenumber{5}
\pubyear{2019}
\copyrightyear{2019}
\history{Received: date; Accepted: date; Published: date}

\usepackage{inconsolata}

\newcommand{\Tabref}[1]{Table \ref{#1}}

\newcommand{\Figref}[1]{Figure \ref{#1}}
\newcommand{\equref}[1]{equation (\ref{#1})}

\newcommand{\cccdot}{\cdot\cdot\cdot}
\newcommand{\Tull}[1]{\T{\ull{#1}}}
\newcommand{\ullWT}[1]{\widetilde{\ull{#1}}}
\newcommand{\TullWT}[1]{\T{\widetilde{\ull{#1}}}}

\newcommand{\sM}{\begin{array}{ccccccccc}}
\newcommand{\eM}{\end{array}}

\let\T\relax
\newcommand{\T}[1]{{#1}^{\sf  T}}

\newcommand{\lb}{\left(}
\newcommand{\rb}{\right)}

\newcommand{\la}{\langle}
\newcommand{\ra}{\rangle}

\newcommand{\sv}{\lb\begin{array}{ccccccccccccccccc}}
\newcommand{\sV}{\begin{bmatrix}}
\newcommand{\eV}{\end{bmatrix}}
\newcommand{\ev}{\end{array}\rb}

\newcommand{\fempty}[1]{{}}

\newcommand{\sty}[1]{\mbox{\boldmath $#1$}}
\newcommand{\styy}[1]{{\mathbb{#1}}}

\newcommand{\fr}{\sty{ r}}

\newcommand{\fx}{\sty{ x}}

\newcommand{\fO}{\sty{ 0}}

\newcommand{\ffR}{\styy{ R}}

\newcommand{\cO}{{\cal O}}
\newcommand{\cP}{{\cal P}}

\newcommand{\scrF}{\mathscr{F}}

\newcommand{\ol}[1]{\overline{#1}}

\renewcommand{\ul}[1]{\underline{#1}}
\newcommand{\ull}[1]{\ul{\ul{#1}}}

\makeatletter
\definecolor{Sblueaa}{cmyk}{1,0.6,0,0}
\definecolor{Sbluea}{cmyk}{1,0.4,0,0}
\definecolor{Sblueb}{cmyk}{0.7,0.2,0,0}
\definecolor{Sbluec}{cmyk}{0.5,0.1,0,0}
\definecolor{Sblued}{cmyk}{0.3,0.05,0,0}
\definecolor{Sbluee}{cmyk}{0.15,0.04,0,0}
\definecolor{Svbluea}{cmyk}{0.9,0.6,0,0}
\definecolor{Svblueb}{cmyk}{0.68,0.4,0,0}
\definecolor{Svbluec}{cmyk}{0.45,0.26,0,0}
\definecolor{Svblued}{cmyk}{0.27,0.12,0,0}
\definecolor{Sblacka}{cmyk}{0.5,0.2,0.2,0.85}
\definecolor{Sblackb}{cmyk}{0.35,0.14,0.14,0.6}
\definecolor{Sblackc}{cmyk}{0.25,0.1,0.1,0.43}
\definecolor{Sblackd}{cmyk}{0.15,0.06,0.06,0.26}
\definecolor{Sblacke}{rgb}{0.827451,0.8509804,0.8627451}
\definecolor{Sred100}{HTML}{EE1C25}
\definecolor{Sorange100}{HTML}{F36F23}
\definecolor{Syellow100}{HTML}{FFDD00}
\definecolor{Spetrol}{HTML}{00AAAD}
\definecolor{Sgreen100}{HTML}{8DC63F}
\definecolor{Spink100}{HTML}{EC008D}
\definecolor{Spurple100}{HTML}{812A91}
\definecolor{Syellow}{cmyk}{0,0.1,1,0}
\definecolor{Sorange}{cmyk}{0,0.7,1,0}
\definecolor{Sred}{cmyk}{0,1,1,0}
\definecolor{Spink}{cmyk}{0,1,0,0}
\definecolor{Spurple}{cmyk}{0.6,1,0,0}
\definecolor{Scyan}{cmyk}{1,0,0.4,0}
\definecolor{Sgreen}{cmyk}{0.5,0,1,0}
\definecolor{Sgreen}{cmyk}{0.5,0,1,0}

\definecolor{uniSgreen}{HTML}{93FF00}
\definecolor{uniSred}{HTML}{FF000B}
\definecolor{uniSpink}{HTML}{FF0098}
\definecolor{uniSorange}{HTML}{FF5D00}
\definecolor{uniScyan}{HTML}{00FBFF}

\colorlet{Sblackf}{Sblacke!70!white}
\colorlet{Sdgreen}{Sgreen!50!black}
\colorlet{Slred}{Sred!40!white}

\usepackage[ruled,lined]{algorithm2e}
\SetAlgoSkip{bigskip}
\LinesNumbered
\g@addto@macro{\@algocf@init}{\SetKwInOut{Input}{Input}}
\g@addto@macro{\@algocf@init}{\SetKwInOut{Output}{Output}}
\g@addto@macro{\@algocf@init}{\SetKwInOut{Requirements}{Requirements\!\!\!\!}}

\definecolor{uniSgray}{RGB}{62, 68, 76}
\colorlet{uniSgray90}{uniSgray!90!white}
\colorlet{uniSgray80}{uniSgray!80!white}
\colorlet{uniSgray70}{uniSgray!70!white}
\colorlet{uniSgray60}{uniSgray!60!white}
\colorlet{uniSgray50}{uniSgray!50!white}
\colorlet{uniSgray40}{uniSgray!40!white}
\colorlet{uniSgray30}{uniSgray!30!white}
\colorlet{uniSgray20}{uniSgray!20!white}
\colorlet{uniSgray10}{uniSgray!10!white}
\definecolor{Scyanlight}{rgb}{ 0.53333,0.87059,0.87451}
\definecolor{uniSblue}{HTML}{004191}
\colorlet{uniSblue80}{uniSblue!80!white}
\colorlet{uniSblue60}{uniSblue!60!white}
\colorlet{uniSblue40}{uniSblue!40!white}
\definecolor{uniSlightblue}{HTML}{00BEFF}
\colorlet{uniSlblue80}{uniSlightblue!80!white}
\colorlet{uniSlblue60}{uniSlightblue!60!white}
\colorlet{uniSlblue40}{uniSlightblue!40!white}
\definecolor{FFgreen}{rgb}{0.635,0.8275,0.1255}
\definecolor{FFdgreen}{rgb}{0.45,0.61,0.09}

\newcommand{\Stilde}{\raise.17ex\hbox{$\scriptstyle\sim$}}

\usepackage{tikz}
\usetikzlibrary{shapes.misc, positioning,arrows,decorations.markings, calc}

\makeatother

\usepackage{multirow}
\usepackage{mathrsfs}
\usepackage{enumitem}

\Title{Data-Driven Microstructure Property Relations}

\Author{Julian Li\ss{}ner$^{1,\dagger,\ddagger}$ and Felix Fritzen$^{2,\dagger,\ddagger}$\orcidB{}}

\AuthorNames{Firstname Lastname, Firstname Lastname and Firstname Lastname}

\address{%
$^{1}$ \quad Institute of Applied Mechanics (CE); lissner@mechbau.uni-stuttgart.de\\
$^{2}$ \quad Institute of Applied Mechanics (CE); fritzen@mechbau.uni-stuttgart.de}

\corres{Correspondence: fritzen@mechbau.uni-stuttgart.de; Tel.: +49-711-685 66283 (F.F.)}

\firstnote{University of Stuttgart, Pfaffenwalding 7, 70569 Stuttgart} 
\secondnote{These authors contributed equally to this work.}

\abstract{An image based prediction of the effective heat conductivity for highly heterogeneous microstructured materials is presented. The synthetic materials under consideration show different inclusion morphology, orientation, volume fraction and topology. The prediction of the effective property is made exclusively based on image data with the main emphasis being put on the 2-point spatial correlation function. This task is implemented using both unsupervised and supervised  machine learning methods. First, a snapshot proper orthogonal decomposition (POD) is used to analyze big sets of random microstructures and thereafter compress significant characteristics of the microstructure into a low-dimensional feature vector. In order to manage the related amount of data and computations, three different incremental snapshot POD methods are proposed. In the second step, the obtained feature vector is used to predict the effective material property by using feed forward neural networks. Numerical examples regarding the incremental basis identification and the prediction accuracy of the approach are presented. A Python code illustrating the application of the surrogate is freely available.}

\keyword{microstructure, property linkage, unsupervised machine learning learning, supervised machine learning, neural network, snapshot proper orthogonal decomposition}

\MSC{74-04, 74A40, 74E30, 74Q05, 74S30}

\usepackage{hyperref}
\begin{document}

\section{Introduction}

In material analysis and design of heterogeneous materials, multiscale modeling can be used for the discovery of microstructured materials with tuned properties for engineering applications. Thereby, it contributes to the improvement of the technical properties, reduces the amount of resources invested into the construction and enhances the reliability of "material modeling"/"the description of material behaviour". However, the discovery of materials with the desired material property, which is characterized by the microstructure of the solid, constitutes a highly challenging inverse problem.

The basis for all multiscale models and simulations is information on the microstructure and on the microscale material behaviour. If at hand, physical experiments can be replaced by -- often costly -- computations in order to determine the material properties by virtual testing~\cite{ghosh1995multiple,dhatt2012finite,leuschner2018}. Separation of structural and microstructural length scales can often be assumed. This enables the use of the representative volume element (RVE) \cite{torquato2013random} equipped with the preferable periodic fluctuation boundary conditions~\citep{jiang2001}. The RVE characterizes the highly heterogeneous material using a single frame (or image) and the (analytical or numerical) computation can be conducted on this frame.

The concurrent simulation of the underlying microstructure (e.g., through nested FE simulations, cf., e.g., \cite{feyel1999,miehe2002}, or e.g. considering microstructure behaviour in the constitutive laws, e.g. \cite{beyerlein2008dislocation}) and of the problem on the structural scale is computationally intractable. In view of the correlation between computational complexity and energy consumption, nested FE simulations should be limited in application. Therefore, efficient methods giving reliable prediction of the material property are an active field of research: POD-driven reduced order models with hyper-reduction (e.g., \cite{ryckelynck2009b,hernandez2014}), with multiple reduced bases spanning also internal variable \cite{fritzen2016b,leuschner2017reduced} and for finite strains (e.g., \cite{yvonnet2007,Kunc2019a}). See also general review articles on the topic such as \cite{kanoute2009,matouvs2017review}.

Supposing that two \textit{similar} images representing microstructured materials are considered, it is natural to expect \textit{similar} effective properties in many physically relevant problems such as elasticity, thermal and electric conduction to mention only two applications. The main task, thus, persists in finding low-dimensional parameterizations of the images that capture the relevant information, use these parameterizations to compress the image information into few numbers and build a surrogate model operating only on the reduced representation. A black-box approach exploiting precomputed data for the construction of the surrogate is the use of established machine learning methods which is also the topic of this paper.

As the no free lunch theorem \cite{wolpert1997no} states, an algorithm can not be arbitrarily fast and arbitrarily accurate at the same time. Hence, there has to be a compromise either in accuracy, computational speed or in versatility. At the cost of generality, i.e. by focusing on subclasses of microstructures, fast \textit{and} accurate models can be deployed while still allowing for considerable variations of the microstructures. This does not mean that these subclasses must be overly confined: for instance, inclusion volume fractions ranging from 20 up to 80\% are considered in this work. Using a limited number of computations performed on relevant microstructure images, machine learned methods can be trained for the subclass under consideration. The sampling of the data, the feature extraction and the training of the ML algorithm constitute the \textit{offline} phase in which the surrogate model is built. Typically, the evaluation of the surrogate can be realized almost in real-time (at least this is the aspired and ambitious objective), thereby enabling previously infeasible applications in microstructure tayloring, interactive user interfaces and computations on mobile devices.

A currently active research for microstructure property linkage is the material knowledge system (MKS) framework \cite{brough2017materials}. Many branches thereof exist, all trying to attain low-dimensional microstructure descriptors from the truncation of selected $n$-point correlation functions. For instance, a PCA of the $n$-point correlation functions of the microstructure is performed and the principal scores are used to in a polynomial regression model in order to predict material properties. The MKS is actively researched for different material structures \cite{paulson2017reduced,gupta2015structure,kalidindi2012computationally}.
For instance, \cite{paulson2017reduced,gupta2015structure} successfully predict the elastic strain and yield stress for the underlying microstructure using the MKS approach, however they confine their focus on either the topological features of the microstructure or a confined range of allowed volume fractions (0-20\%), often held constant in individual studies.

The goal in the present study is to make accurate image based predictions for RVEs spanning large subsets, e.g. in terms of volume fraction, morphological and topological variations, of microstructure materials. 

Similarly to key ideas of the MKS approach, a reduced basis is deployed to reduce the dimensionality of the microstructural features contained in the $n$-point correlation functions. With the sheer amount of samples required, conventional methods fail to capture the key features of the all the microstuctures and we propose three novel incremental reduced basis updates to make the computation possible. Using synthetic microstructures/RVE, the costly training of the reduced basis and of the artificial neural network (e.g., \cite{basheer2000artificial}) becomes feasible, allowing to build a surrogate model for the image-property linkage. The surrogate accepts binarized image representations of bi-phasic materials as inputs. The outputs constitute the effective heat conductivity of the considered material.
 
In Section~2 the microstructure classification and the three different incremental snapshot POD procedures used during feature extraction are presented (unsupervised learning). In Section~3 the use of feed forward artificial neural networks for the processing of the extracted features is discussed. Numerical examples are presented in Section~4 including different inclusion morphologies and an investigation of the relaxation of the microstructure subclass confinement of the procedure by using mixed data sets. A Python code illustrating the application of the surrogate is freely available~\href{\detokenize{https://github.com/J-lissner/img---kappa}}{Github}.

\section{Materials and Methods}
\subsection{Microstructure Classification}
\begin{figure}[b!]
	\includegraphics[width=1.0\textwidth]{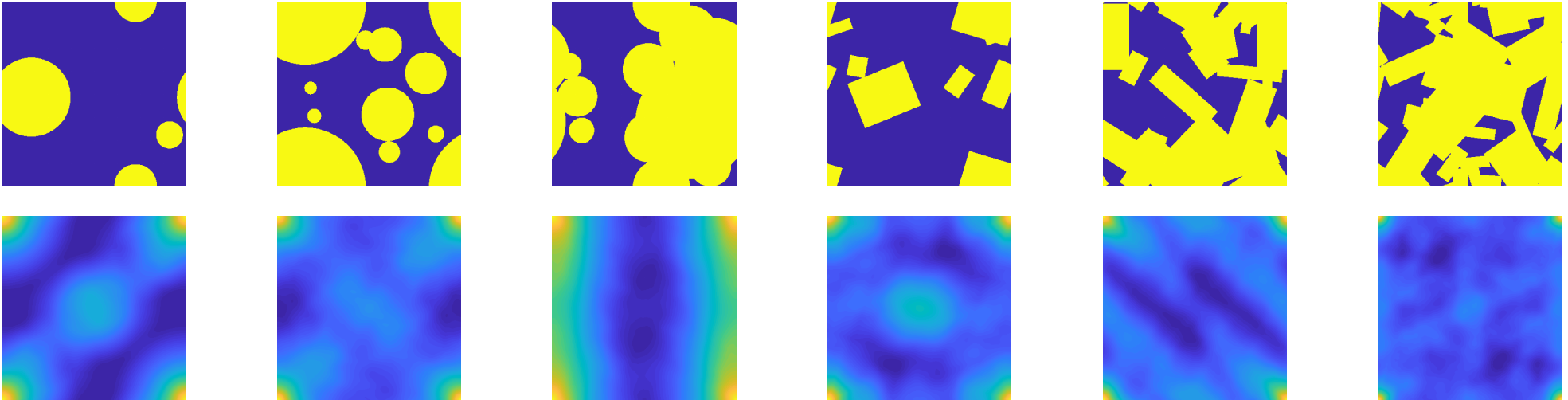}
	\caption{Depicting some exemplary microstructures with their respective 2-point spatial correlation functions $c_2(\fr; b, b)$ below}\label{fig.rves}
\end{figure}
The microstructure is defined by the representative volume element (RVE) \cite{torquato2013random}, which is one periodic frame or image characterizing the heterogeneous material under consideration, see~\Figref{fig.rves} for examples of the microstructure and its 2-point spatial correlation function (see below for its definition). Due to their favorable properties regarding the size of the RVE, periodic fluctuation boundary conditions are used for the computations during the offline phase~\citep{jiang2001}.

The n-point spatial correlation functions represent a widely used mathematical framework for microstructural characterization~\cite{torquato1982microstructure,berryman1985measurement}. Roughly described, the n-point correlation is obtained by placing a polyline consisting of $(n-1)$ nodes defined relative to the first point by vectors $\fr_1, \fr_2, \dots$.

By placing the first point uniformly randomly into the microstructure and computing the mean probability of finding a prescribed sequence of material phases at the nodes of the polyline (including the initial point) denotes the n-point correlation~$c_n(\fr_1, \fr_2, \dots, \fr_{n-1}; m_1, m_2, \dots, m_n)$, where $m_k$ is the material label expected to be found at the $k$th node.


For example, the 1-point spatial correlation function, i.e. the probability of finding phase~$m$ $(m \in \{a,b,\dots\})$, yields the phase volume fraction~$f_m$ of phase $m$. In the present study bi-phasic materials are considered. Here $m=a$ corresponds to the matrix material (drawn blue in \Figref{fig.rves}) and $m=b$ to the inclusion phase (drawn yellow in \Figref{fig.rves}). The trivial relation
\begin{equation}\label{eq.test}
	f_a=1-f_b
\end{equation}
holds. The 2-point spatial correlation function (2PCF) $c_2(\fr;a,b)$ places the vector $\fr$ in each pixel/voxel $\fx$ of the RVE and states the probability of starting in the matrix phase~$a$ and ending in the inclusion phase~$b$. Mathematically we have 
\begin{equation}\label{eq.c2rab}
	c_2( \fr;a,b)=\bigl\langle\, \chi^{(a)}( \fx)\,\chi^{(b)}( \fx+ \fr)\bigl\rangle_x 
\end{equation}
with $\chi^{(m)}$ being the indicator function of phase $m$, $\fr$ the point offset and $\langle \, \bullet\,\rangle_x$ denoting the averaging operator over the RVE. The 2PCF is efficiently computed in Fourier space by making use of the algorithmically sleek fast Fourier transform (FFT)~\cite{cooley1965,FFTW98}:
\begin{align}
    c_2(\fr; a,b) &= \scrF^{-1} \lb  \overline{\scrF(\chi^{(a)})} \odot \scrF(\chi^{(b)}) \rb,
\end{align}
where $\scrF$ and $\scrF^{-1}$ denote the forward and backward FFT, $\ol{\bullet}$ is the complex conjugate and $\odot$ denotes the point-wise multiplication, respectively. For bi-phasic materials the three different two-point functions $c_2(\bullet; a,b)$, $c_2(\bullet; a,a)$, $c_2(\bullet; b,b)$ are related via:
\begin{align}
    c_2(\fr; a,a) &= f_a - c_2(\fr;a,b)\, , &
    c_2(\fr; b,b) &= f_b - c_2(\fr;a,b) \, .
\end{align}
In view of computational considerations this redundancy can be exploit. Some key characteristics of the non-negative 2PCF are
\begin{align}
 c_2(\fO; a,a) &= f_a = \underset{\fr\in\varOmega}{\text{max}} \; c_2(\fr; a,a), \\
 c_2(\fO; b,b) &= f_b = \underset{\fr\in\varOmega}{\text{max}} \; c_2(\fr; b,b), \\
 c_2(\fO; a,b) &= 0, \\
 c_2(\fr; a,b) &= c_2(\fr; b,a) = c_2(-\fr; a,b), \\
 \la c_2(\fx; m,m) \ra_x & = f_m^2 && (m=a, b)\,.
\end{align}
In addition to that, a key property of the 2PCF is its invariance with respect to translations of the periodic microstructure. This property is of essential importance when it comes to the comparison of several images under consideration, i.e. during the evaluation of similarities within images.

Examples of $c_2(\fr; b,b)$ (referred to also as auto-correlation of the inclusion phase) are depicted by the lower set of images in \Figref{fig.rves}. By the metric of vision, the following characteristics can be observed:
\begin{itemize}
 \item the maximum of $c_2(\fr; b,b)$ occurs at the corners of the domain (corresponding to $\fr=\fO$);
 \item preferred directions of the inclusion placement and/or orientation correspond to laminate-like images (best seen in the third microstructure from the left);
 \item the domain around $\fr=\fO$ partially reflects the average inclusion shape;
 \item some similarities are found, particularly with respect to shape of the 2PCF at the corners and in the center.
\end{itemize}
These observations hint at the existence of a low-dimensional parametrization of relevant microstructural features. In the following this property is exploited by using a snapshot proper orthogonal decomposition (snapshot POD) in order to capture reoccurent patterns of the 2PCF. By working on the two-point function the afore-mentioned elimination of possible translations of the images is an important feature.

The influence of higher order spatial correlation functions has been investigated in the literature \citep[e.g.,][]{torquato1982microstructure,fast2011formulation}. These considerations often yield minor gains relative to the additional computations and the increased dimensionality\footnote{For instance, the 3PCF takes to vectors $\fr_1,\fr_2\in\varOmega$ as inputs. Hence, the full 3PCF is basically inaccessible in practice but only after \textit{major} truncation.}. Although it has been demonstrated that the two point function does not suffice to uniquely describe the microstructure in periodic domains \cite{fullwood2008microstructure}, there is evidence that the level of microstructural ambiguity for identical 2PCF can be considered low. Therefor, only the n-point correlation functions up to second order are accounted for in the present study.

\subsection{Unsupervised Learning via Snapshot Proper Orthogonal Decomposition}
\label{sec.pod}
The snapshot POD \cite{sirovich1987} can be used to construct a reduced basis (RB) \cite{LIANG2002527,camphouse2008snapshot,quarteroni:2016} that provides an optimal subspace for approximating a given snapshot matrix $\ull S \in \styy R^{n\times n_s}$. The matrix~$\ull{S}$ consists of $n_s$ individual snapshots $\ul s_i\in \styy R^{n}$ with the size $n$ being the dimension of the discrete representation of the unreduced field information. In the case of the 2PCF $n$~denotes the total number of pixels within the RVE, i.e. the discrete two-dimensional 2PCF (representing image data) is recast into vector format for further processing ($\,\ul c^0_2(m,m) \in \ffR^n\,$). 
In the present study, the constructed RB is used for information compression, i.e. for the extraction of relevant microstructural features from the image data. The reduced basis $\ull B \in \ffR^{n\times N}$ retains the $N$ most salient features of the data contained in~$\ull{S}$ in a few eigenmodes represented by the orthonormal columns of $\ull{B}$.

The actual snapshot data contained in $\ull{S}$ is thus constructed from the discrete 2-point function data $\ul{s}^0_i\in\ffR^n$ via
\begin{align}\label{eq.s_scaling}
\ul s_i=\ul c^0_{2_i}(b,b)-f_b^2 \,\ul{1}=\ul c^0_{2_i}(a,a)-f_a^2 \,\ul{1}\,.
\end{align}
where $\ul{1}\in\ffR^n$ is a vector containing ones at all entries.
The reduced basis is computed under the premise to minimize the overall relative projection error
\begin{equation}
	\mathcal P_{\delta}=\frac{||\ull{S} -\ull B\,\Tull B\,\ull{S}||_F}{||\ull{S}||_F}
\end{equation}
with respect to the Frobenius norm $\Vert\bullet\Vert_F$.
The RB can be constructed with multiple methods, e.g. with the snapshot correlation matrix $\ull C_S$ and its eigenvalue decomposition, which is given by 
\begin{equation}
	\ull C_S=\Tull S\,\ull S=\ull V\,\ull{\Theta}\,\Tull V\,.
\end{equation}
The following properties of the sorted eigenvalue decomposition hold:
\begin{align}
	\T{\ull V}\,\ull V&=\ull I\qquad \styy R^{n_s\times n_s}\,, &
	\Theta_{ij} &=\theta_i\delta_{ij}\,, &
	\theta_1 &\geq\theta_2\geq...\geq \theta_{n_s}\geq 0\,,
\end{align}
and $\delta_{ij}$ denotes the Kronecker delta. The dimension of the reduced basis is determined by the POD threshold, i.e. the truncation criterion is given by
\begin{align}
	\delta_N=\sqrt{\frac{\sum_{j=N+1}^{n_s}\theta_j}{\sum_{i=1}^{n_s}\theta_i}}=\sqrt{\frac{\sum_{j=N+1}^{n_s}\theta_j}{||\ull S||^2_F}}
	= \sqrt{ \Vert \ull{S} \Vert_F^2 - \sum_{j=1}^N \theta_j } \stackrel{!}{\leq} \varepsilon\,, \label{eq:pod:threshold1}
\end{align}
where $\varepsilon>0$ is a given tolerance denoting the admissible approximation error. Then, the reduced basis is computed via 
\begin{equation}\label{eq.csrb}
	\ull B =\ull S\,\ullWT{V}\,\ullWT{\Theta}^{-\frac12}
\end{equation}
after truncation of the eigenvalue- and eigenvector matrices to reduced dimension $N$ represented with $\ullWT{\Theta} \in \styy R^{N\times N}$ and $\ullWT V \in \styy R^{n\times N}$, respectively. The sorting of the eigenvalues with their corresponding eigenvectors leads to the property that the least reoccurent information given in $\ull S$ is omitted. Hence, the first eigenmode in $\ull B$ has the most dominant pattern, the second eigenmode the second most etc.\\
The properties of the reduced basis computed with the snapshot correlation matrix remain the same as for the singular value decomposition (SVD) introduced below.

The SVD \cite{klema1980singular} of the snapshot matrix is given by
\begin{equation}
	\ull S=\ull U\,\ull {\Sigma}\, \T{\ull{W}}
\end{equation}
with the following properties (asserting $n_s\geq n$)
\begin{align}
    \ull{U}\in\ffR^{n\times n_s}:\ 
	\T{\ull{ U}}\,\ull{ U} &= \ull{ I}\,, &
	\ull{W}\in\ffR^{n_s\times n_s}:\ \T{\ull{ W}}\,\ull{ W} &= \ull{ I}\,, &
	\ull{\Sigma}\in\ffR^{n_s\times n_s}:\ \ull{\Sigma} &= \text{diag}(\sigma_i)
\end{align}
and the sorted non-negative singular values $\sigma_i$ such that $\sigma_1\geq \sigma_2\geq\cccdot\geq \sigma_{n_s}\geq 0$. The criterion for determining the reduced dimension~$N$ matching \eqref{eq:pod:threshold1} takes the form
\begin{align}
	\delta_N &= \sqrt{\frac{\sum_{j=N+1}^{n_s}\sigma^2_j}{\Vert \ull{S} \Vert_F^2}} \,
    = \sqrt{\frac{\sum_{j=N+1}^{n_s}\sigma^2_j}{\sum_{i=1}^{n_s}\sigma^2_i}} \,
    = \sqrt{ \Vert \ull{S} \Vert_F^2 - \sum_{j=1}^N \sigma_j^2 } \overset{!}{\leq} \varepsilon.
\end{align}
Then, the reduced basis is given by truncation of the columns of $\ull{U}\to\ullWT{U}\in\ffR^{n \times N}$
\begin{equation}
	\ull B=\ullWT U.
\end{equation}
More specifically, the left subspace associated with the leading singular values is the RB. Both introduced methods yield the exact same result for the same snapshot matrix $\ull S$.

\subsection{Incremental Generation of the Reduced Basis $\ull B$}\label{sec.inc.RB}
The RB is deployed in order to compress the information contained in $n_s$~snapshots into an $N$-dimensional set of eigenmodes stored in the columns of $\ull B\in\ffR^{n\times N}$, where $N\ll n_s$ is asserted. Since the RB is computed with the snapshot matrix alone, the information contained in $\ull S$ needs to contain data representing the relevant microstructure range, i.e. covering the parameter range used in the generation of the synthetic materials, in order for $\ull B$ to be representative for the problem under consideration.

In the case of microstructural images containing $n$ pixels, a ludicrous amount of $2^n$ states could theoretically be considered when allowing for fully arbitrary microstructures. When limiting attention to certain microstructure classes, then less information is required. Still, thousands of snapshots are usually required, at least. In the following, attention is limited to synthetic materials generated using random sequential adsorption of morphological prototypes with variable size, orientation, aspect ratio, overlap and phase volume fraction. Due to the high variability of such microstructures (see, e.g., \Figref{fig.rves}), a large number of snapshots is required that can usually not be stored in memory, i.e. a monolithic snapshot matrix~$\ull{S}$ is not available. Although attention is limited to two-dimensional model problems in this study, the problem aggravates \textit{considerably} for three-dimensional images which imply technical challenges of various sort (storage, processing time, data management, \dots).\\

\begin{figure}[h!]
    \centering
	\includegraphics[scale=1]{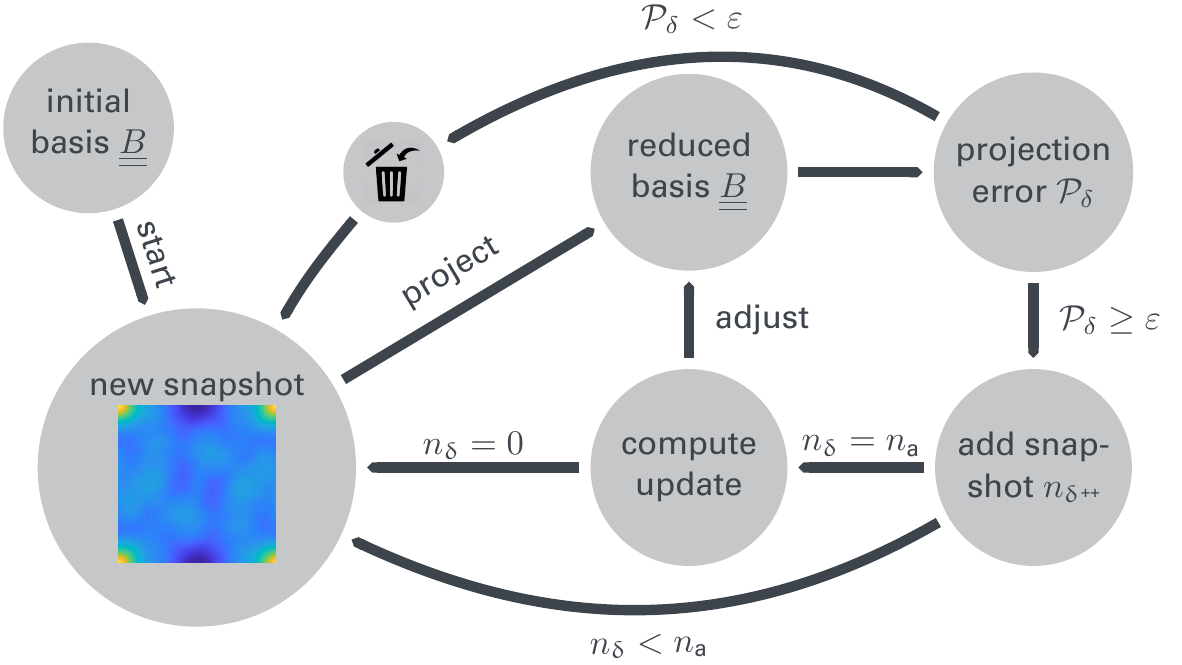}
	\caption{Graphical overview of the incremental update of the reduced basis}\label{fig.it_b}
\end{figure}

In order to be able to generate a rich RB accounting for largely varying microstructural classes, the incremental basis generation represents a core concept within the present work. It enables the RB generation based on a sequence of input snapshots but without the need to store previously considered data except for the current RB.
Three different methods are proposed, two of which rely on approximations of the snapshot correlation matrix~$\ull{C}_S$, and one of which relies on the SVD of an approximate snapshot matrix. The general incremental scheme depicted in \Figref{fig.it_b} remains the same for all the procedures, i.e. the only difference is found during the step labeled \textit{adjust}.

The algorithm is initialized  by a small sized set of initial snapshots of the shifted 2-point correlation function (see Section~\ref{sec.pod}). Further, the algorithmic variables $n_\delta=0$ and $\ull{\Delta S}=\ull{0}$ are set. The initial RB is computed classically using either the correlation matrix or the SVD (see previous section for details). After computation of the RB, the snapshots are stored neither in memory nor on a hard drive. The algorithm then takes input snapshots in the order of appearance. For each new snapshot~$\ul{s}_i$ the relative projection error with respect to the current RB is computed:
\begin{equation}\label{eq.pdelta}
	\mathcal P_{\delta}=\frac{||\ul s_i-\,\ull B\,\Tull B\,\ul s_i||_F}{||\ul s_i||_F}\,.
\end{equation}
If $\cP_\delta$ is greater than the tolerance $\varepsilon > 0$ the snapshot is considered as inappropriately represented by the existing RB. Consequently, $\ul{s}_i$ is appended to a buffer $\ull{\Delta S}$ containing candidates for the next basis enrichment and the counter $n_\delta$ is incremented. Once the buffer contains a critical number of $n_{\rm a}$ elements the actual enrichment is triggered and the buffer is emptied thereafter. Thereby the computational overhead is reduced. The three different update procedures are described later on in detail. The procedure is continued until $n_{\rm c}>0$ consecutive snapshots were found to be approximated up to the relative tolerance~$\varepsilon$. Then the basis is considered as converged for the microstructure class under consideration.

In the following three methods for the update procedure are described. Formally, the update of an existing basis $\ull B$ with a block of snapshots contained in the buffer $\ull{\Delta S}$ is sought-after. The new basis is required to remain orthonormal.

\subsubsection{Method A: Append Eigenmodes to $\ull B$}
A trivial enrichment strategy is given in terms of appending new modes to the existing basis while preserving orthonormality of the basis. Therefore, the projection of $\ull{\Delta S}$ onto the existing RB is subtracted in a first step:
\begin{align}
 \ull{\Delta \widehat{S}} & = \ull{\Delta S} - \ull{B} \, \T{\ull{B}} \ull{\Delta S}. 
\end{align}
It is readily seen that $\ull{\Delta\widehat{S}}$ is orthogonal to $\ull{B}$. Then the correlation matrix of the additional data and its eigen-decomposition are computed according to
\begin{align}
 \ull{\Delta C} &= \T{\ull{\Delta\widehat{S}}} \ull{\Delta\widehat{S}} = \ull{V}\,\ull{\Theta}\,\T{\ull{V}} .
\end{align}
Eventually, the enrichment is given through the truncated matrices $\ullWT{V}$ and $\ullWT{\Theta}$:
\begin{align}
 \ull{\Delta B} &= \ull{\Delta \widehat{S}} \, \ullWT{V} \, \ullWT{\Theta}^{-\tfrac{1}{2}}.
\end{align}
The new basis is then obtained by
\begin{align}
 \ull{B} & \leftarrow \sV \ull{B} & \ull{\Delta B} \eV.
\end{align}
Method A simply adds modes generated from the projection residual $\ull{\Delta\widehat{S}}$ in a decoupled way, i.e. the existing basis is not modified. In order to compute the basis update, only the existing RB $\ull B$ and the temporarily stored snapshots $\ull{\Delta S}$ are required.
\paragraph*{Remarks on Method A}
\begin{enumerate}[label=\textbf{A.\arabic*{}}]
 \item  The truncation parameter~$\epsilon$ must be chosen carefully such that
\begin{align}
 \frac{ \Vert \ull{\Delta\widehat{S}} - \ull{\Delta B} \, \T{\ull{\Delta B}} \ull{\Delta\widehat{S}} \Vert_F }{\Vert \ull{\Delta S} \Vert_F} & \leq \epsilon.
\end{align}
In particular, the normalization with respect to the original data prior to projection onto the existing RB must be taken.
\item By appending orthonormal modes to the existing basis it is \textit{a priori} guaranteed that the accuracy of previously considered snapshots cannot worsen, i.e. an upper bound for the relative projection error of all snapshots considered until termination of the algorithm is given by the truncation parameter~$\epsilon$ and $n_{\rm a}$:
\begin{align}
 \text{max} \, \frac{\vert \ul{s}_i - \ull{B}\, \T{\ull{B}} \ul{s}_i \vert}{\vert s_i \vert} & \leq \sqrt{n_{\mathrm{a}}}\, \epsilon .
\end{align}
\end{enumerate}

\subsubsection{Method B: Approximate Reconstruction of the Snapshot Correlation Matrix}
The goal of this iterative update scheme is the accurate approximation of the new correlation matrix
\begin{equation}\label{eq.csbig}
	\ull C=\begin{bmatrix}
		\T{\ull{S}} \ull{S} & \T{\ull S}\,\ull{\Delta S} \\[2mm]
		\T{\ull{\Delta S}}\,\ull{S} & \T{\ull{\Delta S}}\,\ull{\Delta S}
		\end{bmatrix}
        = \begin{bmatrix}
		\ull{C}_0 & \T{\ull S}\,\ull{\Delta S} \\[2mm]
		\T{\ull{\Delta S}}\,\ull{S} & \T{\ull{\Delta S}}\,\ull{\Delta S}
		\end{bmatrix} \, .
\end{equation}
Here $\ull{S}$ denotes all snapshots considered in the RB so far and $\ull{\Delta S}$ contains the candidate snapshots. However, the previously used snapshots formally written as $\ull{S}$ are no longer available since they can not be stored due to storage limitations.
Using the previously computed matrices $\ull B,\ullWT V, \ullWT{\Theta}$ the following approximations are available:
\begin{align}
		\Tull S\,\ull S & =\ull C_0 \approx \widetilde{\ull C}_0=\ullWT V \,\ullWT{\Theta} \,\TullWT V \, , &
		\ull B &= \ull S\,\widetilde{\ull V}\,\widetilde{\ull{\Theta}}^{-\frac12} \, , &
		\ull S &\approx \ull B\,\T{\ull B}\,\ull S \, ,
\end{align}
where the accuracy of the approximation is governed by the truncation threshold $\delta_N$. Using these approximations and using the property of the eigenvalue decomposition, the snapshot matrix $\ull S $ can be approximated by 
\begin{equation}
	\ull S \approx \ull B \,\ullWT{\Theta}^{\frac12}\,\TullWT V.
\end{equation}
Note that $\ull{B}\in\ffR^{n\times N}$ is stored anyway, $\ullWT{\Theta}\in\ffR^{N\times N}$ is diagonal and $\ullWT{V}\in\ffR^{n \times N}$ is of manageable size.
The snapshot correlation matrix $\ull C$ that considers the additional snapshots can be approximated as 
\begin{equation} 
	\ull C\approx \begin{bmatrix}
		\widetilde{\ull C}_0 \quad& \widetilde{\ull V}\,\widetilde{\ull{\Theta}}^{\frac12}\,\T{\ull B}\,\ull{\Delta S}\\[2mm]
		\T{\ull{\Delta S}}\,\ull B\,\widetilde{\ull{\Theta}}^{\frac12}\,\T{\widetilde{\ull V}} & \T{\ull{\Delta S}}\,\ull{\Delta S}
		\end{bmatrix} 
	=\underbrace{\begin{bmatrix} \ullWT V&\ull 0 \\[2mm] \ull 0 & \ull I \end{bmatrix}}_{\ull V_{\ast}} \,
		\underbrace{\begin{bmatrix} \ullWT{\Theta} & \ullWT{\Theta}^{\frac12}\,\Tull B\,\ull{\Delta S} \\[2mm]
					\text{sym.} & \Tull{\Delta S}\,\ull {\Delta S}
			\end{bmatrix}}_{\ull C_{1}} \,
	\underbrace{\begin{bmatrix} \TullWT V&\ull 0 \\[2mm] \ull 0 & \ull I \end{bmatrix}}_{\Tull V_{\ast}}\,.
\end{equation}
In order to compute the updated basis, the inexpensive eigenvalue decomposition of $\ull C_{1}\in \styy R^{(N+n_{\rm a})\times (N+n_{\rm a})}$ is computed
\begin{equation}
	\ull C_{1}=\ull V_1 \,\ull{\Theta}_1 \,\Tull V_1\,.
\end{equation}
Analogously to the previous RB computation in \equref{eq.csrb}, the adjusted and enriched basis is computed by 
\begin{equation}
	\ull B=\Big[ \ull S \quad \ull{\Delta S} \Big] \,\ullWT V\,\ullWT{\Theta}^{-\frac12}
		\approx \Big[ \ull B \,\ullWT{\Theta}^{\frac12}\,\TullWT V \quad \ull{\Delta S} \Big] \,\underbrace{\ull V_{\ast}\,\ullWT{V}_1 }_{\ullWT{W}}\,\ullWT{\Theta}_1^{-\frac12} .
\end{equation}
To update the RB the truncated eigenvector matrix ($\ull B\,,\ullWT V \leftarrow \ullWT W \in\styy R^{n\times N}$) need to be stored as well as the diagonal eigenvalue matrix $\ullWT{\Theta}$.

\paragraph*{Remarks on Method B}
\begin{enumerate}[label=\textbf{B.\arabic*{}}]
 \item  The existing RB is not preserved but it is updated using the newly available information. Thereby, the accuracy of the RB for the approximation of the previous snapshots is not guaranteed \textit{a priori}. However, numerical experiments have shown no increase in the approximation errors of previously well-approximated snapshots.
 \item  In contrast to Method A the dimension of the RB can remain constant, i.e. a mere adjustment of existing modes is possible. The average number of added modes per enrichment is well below that of Method A.
 \item The additional storage requirements are tolerable and the additional computations are of low algorithmic complexity.
\end{enumerate}

\subsubsection{Method C: Incremental SVD}
Method C is closely related to Method B. However, instead of building on the use of the correlation matrix, it relies on the use of an updated SVD, i.e. an approximate truncated SVD is sought-after:
\begin{equation}\label{eq.svd_full}
	\text{trunc\,svd}\sv \Big[\ull S\quad \ull{\Delta S} \Big]\ev \approx \ull B\,\ull{\Sigma}\,\Tull W\,.
\end{equation}
Since the original snapshot matrix $\ull S$ can not be stored, only an approximation of the actual truncated SVD in \eqref{eq.svd_full} can be computed. Methods to compute an incremental SVD are introduced in \cite{gu1994stable,FAREED20181942}, with the latter referring to Brand's incremental algorithm \cite{horn1985matrix} which is used in the present study with minor modifications.
With the previously computed basis $\ull B$ at hand, the approximation of $\ull S$ is known
\begin{equation}\label{eq.svd_s}
	\ull S\approx\ull B\,\ull{\Sigma}\,\T{\ull W}\,.
\end{equation}
Introducing the projection residual $\ull{\Delta \widehat{S}}$ of the enrichment snapshots $\ull {\Delta S}$ together with its SVD 
\begin{equation}\label{eq.orthoproj_svd}
 	\ull{\Delta \widehat{S}}=\ull{\Delta S}-\ull B\,\T{\ull B}\,\ull{\Delta S} \quad
	=\,\ull U_S\,\ull{\Sigma}_S\,\Tull W_S \,,
\end{equation}
and using \eqref{eq.svd_full}, \eqref{eq.svd_s} and \eqref{eq.orthoproj_svd} the full snapshot matrix can be approximated after some algebra by
\begin{equation}\label{eq.add_S}
	\big[\ull S \quad \ull{\Delta S}\big]\approx \big[\ull B\,\ull{\Sigma}\,\T{\ull W}\quad\ull {\Delta S}\big]
	= \big[\ull B\quad \ull U_S\big]
	\underbrace{
	\begin{bmatrix}
		\ull{\Sigma} & \T{\ull B}\,\ull{\Delta S}\\[1.5mm]	\ull 0&\ull{\Sigma}_S\,\Tull W_S
	\end{bmatrix}
		}_{\ull{\Gamma}}
	\T{\begin{bmatrix}
		\ull W & \ull 0 \,\\[1.5mm] \ull 0 &\ull I
	\end{bmatrix}}\,.
\end{equation}
Considering the SVD of $\ull{\Gamma} =\ull U_{\Gamma}\,\ull{\Sigma}_{\Gamma}\,\T{\ull W}_{\Gamma}\in \styy R^{(N+n_{\rm a})\times(N+n_{\rm a})}$, which is inexpensive due to the sparsity of $\ull{\Gamma}$, approximation \eqref{eq.add_S} is further rewritten as
\begin{equation}\label{eq.svd_hidden}
	\big[\ull S\quad\ull{\Delta S}\big]
	\approx \Big(\underbrace{\big[\ull B\quad\ull U_S\big]\,\ull U_{\Gamma}\,}_{\ull U_{\ast}} \Big)\,
	\underbrace{\ull{\Sigma}_{\Gamma}}_{\ull{\Sigma}_{\ast}}
	\bigg( \underbrace{	\begin{bmatrix}
		\ull W & \ull 0 \\[1.5mm] \ull 0 &\ull I
	\end{bmatrix}
	\,\ull W_{\Gamma}}_{\ull W_{\ast}} \T{\bigg)}\,.
\end{equation} 
It is easily shown that the matrices $\ull U_{\ast}$ and $\ull W_{\ast}$ are column-orthogonal and that $\ull{\Sigma}_{\ast}$ is diagonal and non-negative. Therefore, the three matrices constitute an approximate SVD of the enlarged snapshot matrix at low computational expense. This implies the following updates after each enrichment step
\begin{equation}\label{eq.svd_updt}
	\ull B \leftarrow \Big[\ull B\quad\ull U_S\,\Big]\,\ull U_{\Gamma}
	\hspace{2.5cm} \ull{\Sigma}\leftarrow\ullWT{\Sigma}_{\Gamma}
	\hspace{2.5cm} \ull W\leftarrow 
		\begin{bmatrix}
			\ull W & \ull 0 \\ \ull 0 &\ull I
		\end{bmatrix}
		\ull W_{\Gamma} 
\end{equation}
after truncation of $\ull B$, where the truncation criteria needs to ensure that $\ull B$ does not decrease in size. To compute the enrichment of the RB, $\ull B\in \styy R^{n\times N}$ and the sparse singular values $\ull{\Sigma} \in \styy R^{N\times N}$ after truncation need to be stored.

\paragraph*{Remarks on Method C}
\begin{enumerate}[label=\textbf{C.\arabic*{}}]
 \item  As highlighted for Method B in remark \textbf{B.1}, the existing RB is not preserved but adjusted by considering the newly added information. This ensures the accuracy of the approximation error of previously well-approximated snapshots, but \textit{a priori} guarantees regarding the subset approximation accuracies cannot be made.
 \item  In contrast to Method A the dimension of the RB can remain constant, i.e. a mere adjustment of existing modes is possible. The average number of added modes per enrichment is well below that of Method A.
 \item Each update step in \equref{eq.svd_updt} is computed separately and, consequently, storing $\ull W$ is not required since only the RB $\ull B$ is of interest.
 \item The diagonal matrix $\ull {\Sigma}$ has low storage requirements corresponding to that of a vector in $\ffR^N$.
\end{enumerate}

\section{Supervised learning using Feed Forward Neural Network}\label{sec.nn}
During the supervised learning phase, the machine is provided with data sets consisting of inputs and the related outputs. Hence, the supervised learning phase tries to learn a function relating inputs to outputs without or with limited prior knowledge of the structure of the unknown mapping. Artificial neural networks (ANN) are a powerful machine learning tool which have gained wide popularity in the recent decades due to the surge in computational power \cite{widrow199030,basheer2000artificial}. 

The functionality of the ANN is inspired by the (human) brain, propagating a signal (input) through multiple neurons where it is lastly transformed into an action (output). Various types of neural networks have been invented, e.g. feed forward, recurrent, convolutional, being applicable to almost any field of interest \cite{kimoto1990stock,sundermeyer2012lstm,simonyan2014very,angermueller2016deep}.
\begin{figure}[!h]
	\centering
	\includegraphics[width=1.0\textwidth]{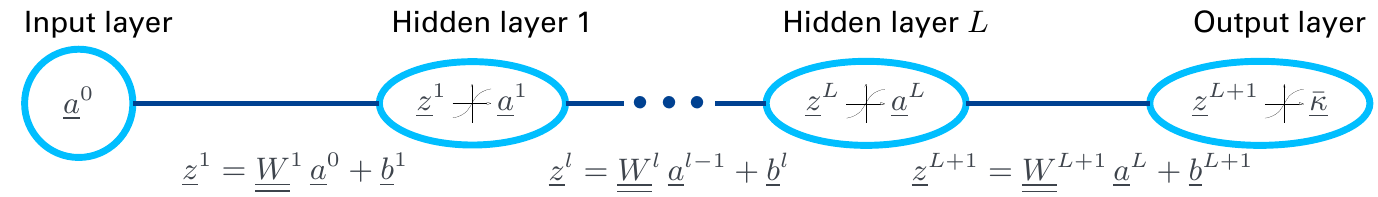}
	\caption{The basic functionality of a dense feed forward neural network is depicted in simplified form.}\label{fig.nn}
\end{figure}
In the present study a regression model from the input, i.e. the feature vector $\ul{\xi}$ which is derived with the converged basis $\ull B$, to the output, i.e. the effective heat conduction tensor $\bar{\ull{\kappa}}$, is deployed with a dense feed forward ANN. 

In a dense feed forward ANN (\Figref{fig.nn}) a signal is propagated through the hidden layers where every output of the previous layer $\ul a^{l-1}$ affects the activation $\ul z^l$ of the current layer $l \, (l=1,\dots,L+1)$. The activation of each layer gets wrapped into an activation function $f$ where the output of each neuron in the layers is computed, i.e. $\ul a^l=f(\ul z^l)$. Note that matrix/vector notation is used, where each entry in the vectors denotes one neuron in the respective layer.

The basic learning algorithm/optimizer usually employed for a feed forward ANN is the back propagation algorithm \cite{hecht1992theory} and modifications thereof.
The learning of the network consists in the numerical identification of randomly initialized and unknown weights $\ull W^l$ and biases $\ul b^l$ that minimizes a given cost function. The latter gives an indication of the quality of the ANN prediction. For instance, the gradient back propagation computes gradients of the cost function with respect to the weights and biases, in order to compute suitable corrections for these parameters.

The learning itself is an iterative procedure in which the training data is cycled multiple times through the ANN (one run called an \textit{epoch}). In each epoch the internal parameters are updated with the aim of improving the mapping from the input to the output. 
The optimization problem itself is (usually) high-dimensional. In most situations it is not well-posed and local minima and maxima can hinder convergence. Therefor, multiple random instantations of the network parameters are usually required to assure that a good set of parameters is found, even if the network layout remains unaltered. 

Notably, the training requires a substantial input data set as inputs. It is important to note that the training of the ANN usually results in a parameter set that is able to approximate the training data with high accuracy under the given meta-parameters describing the network architecture (number of layers, number of neurons per layer, type of activation function). However, the approximation quality of the ANN may be different for other queries. Thus, it is important to validate the generality of the discovered mapping for the underlying problem setting. Therefor an additional validation data set is introduced, where only the evaluation of the cost function is tracked during the training. Generally when overfitting\footnote{Overfitting relates to the fact that a subset of the data is nicely matched but small variations in the inputs can lead to substantial loss in accuracy, similar to oscillating higher-order polynomial interpolation functions.} occurs, the errors for the validation set increase whereas the errors of the training set decrease. The training should be halted if such a scenario is detected.

Since the choice of activation function as well as the number of hidden layers and the number of neurons within the individual layers are arbitrary (describing the ANN architecture), these meta-parameters should be tailored specifically for the desired mapping. Finding the best neural network architecture is not straight-forward and usually relies on intuition, experience and many numerical experiments. In order to find a well suited ANN, various (random) realizations of each tested ANN architecture need to be computed, before a decision regarding the optimal layout can be made.

In the present study the ANN training is performed using \href{https://www.tensorflow.org}{Tensorflow} in Python \cite{abadi2016tensorflow}. Tensorflow is an open source project by the Google team, providing highly efficient algorithms for ANN implementation. The Adam \cite{kingma2014adam} optimizer, which is a modification of the gradient back propagation, has been deployed for the learning.

\section{Results}\label{sec.results}
\subsection{Generation of Synthetic Microstructures}
All of the used synthetic microstructures have been generated by a random sequential adsortion algorithm with some examples shown in \Figref{fig.rves}. Two morphological prototypes were used: spheres and rectangles. The parameters used to instantiate the generation of a new microstructure were modeled as uniformly distributed variables:
\begin{enumerate}[label=\textbf{M.\arabic*}]
 \item the phase volume fraction $f_b$ of the inclusions (0.2-0.8);
 \item the size of each inclusion (0.0--1.0);
 \item for rectangles: the orientation (0--$\pi$) and the aspect ration (1.0--10.0);
 \item the admissible relative overlap $\varrho$ for each inclusion (0.0--1.0).
\end{enumerate}
For $\varrho=0$ and the spherical inclusion, a boolean model of hard spheres is obtained. Setting $\varrho=1$ induces a boolean model without placement restrictions, i.e. new inclusions can be placed independent of the existing ones. The generated microstructures were stored as images with resolution 400$\times$400.
After the generation of the RVE, the 2-point spatial correlation function was computed for the RVE. This was then shifted, see Section~\ref{sec.pod}, and used as a snapshot $\ul s_i$ for the identification of the reduced basis.

\begin{figure}[h!]
	\includegraphics[width=1.0\textwidth]{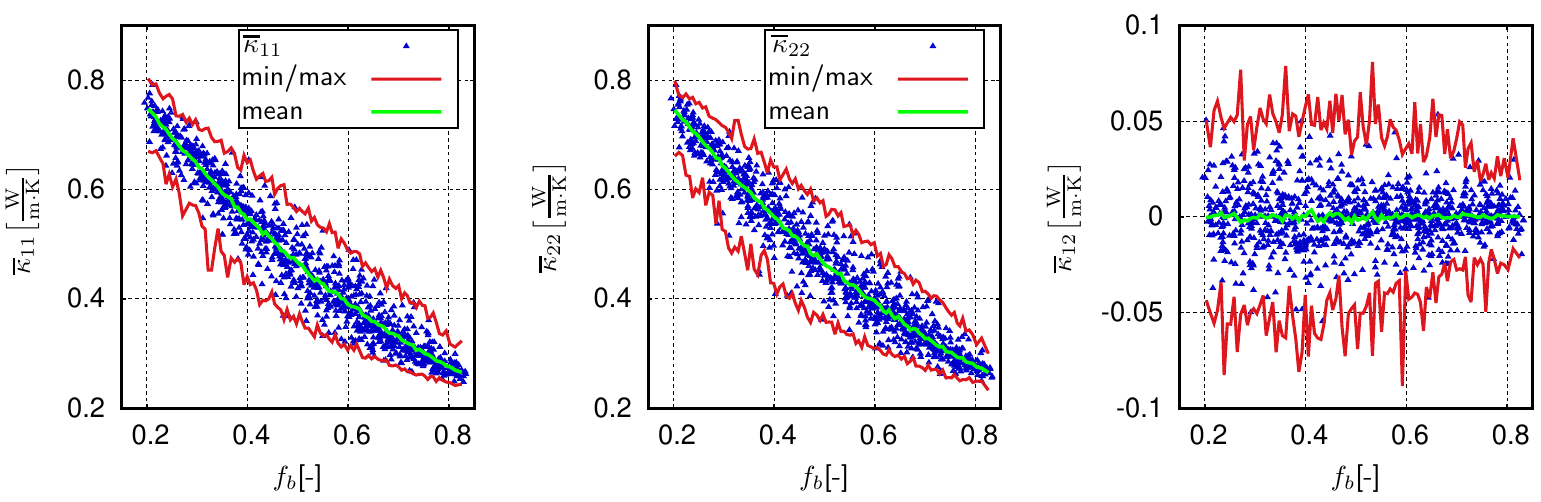}
	\caption{The range of each $\ol{\ull{\kappa}}$ entry computed with 15000 microstructures is of the mixed set. Only 1000 discrete values are shown in each plot.}\label{fig.rangekappa}
\end{figure}

Additionally, a smaller random set of RVEs used for the supervised learning phase was simulated using the recent Fourier-based solver FANS \cite{leuschner2018} in order to compute the effective heat conduction tensor~$\bar{\ull{\kappa}}$. The heat conductivity of the matrix and of the inclusion phase are prescribed as
\begin{equation}\label{eq.kappa}
	\kappa_a=1.0 \big[\frac{\rm W}{\text m\cdot\text K}\big]\,,\hspace{2.5cm}
	\kappa_b= \frac{\kappa_a}{R} \,  \big[\frac{\rm W}{\text m\cdot\text K}\big] \, .
\end{equation}
Here $R>0$ denotes the material contrast. In the present study, $R=5$ was considered, i.e. the matrix of the microstructure has a five times higher conductivity than the inclusions. These values can be seen as typical values for metal ceramic composites (\Figref{fig.rangekappa}).

An inverse phase contrast has exemplarily been studied, i.e. inclusions with $\kappa_b=1$ and $\kappa_a=\frac{\kappa_b}5$ (corresponding to $R=\tfrac{1}{5}$) has also been investigated. Qualitatively, the results for the inverse phase contrast did not show any new findings or qualitative differences. Therefor, the following results focus on $R=5$, corresponding to rather insulating inclusions.

The symmetric tensor $\bar{\ull{\kappa}}$ can be represented as a three-dimensional vector $\bar{\ul{\kappa}}$ using the normalized Voigt notation
\begin{equation}
	\ull{\bar{\kappa}}=\begin{bmatrix} \bar{\kappa}_{11}\,&\,\bar{\kappa}_{12}\\\bar{\kappa}_{21}&\bar{\kappa}_{22}\end{bmatrix} 
	\qquad\to\qquad \ul{\bar{\kappa}}_{\rm V}=\begin{bmatrix} \bar{\kappa}_{11}\\ \bar{\kappa}_{22}\\\sqrt2 \,\bar{\kappa}_{12} \end{bmatrix}\,.
\end{equation} 
For the supervised learning of the ANNs (see Section~\ref{sec.nn}),  multiple files each containing 1500 data sets for different inclusion morphologies were generated (circle only; rectangle only; mixed). Each data set contains the image of the microstructure, the respective autocorrelation of the inclusion phase~$c_2(\bullet; b,b)$ and the effective heat conductivity~$\bar{\ul{\kappa}}_{\mathrm{V}}$.


\subsection{Unsupervised Learning}
\label{sec.res_rb}

First, the reduced basis is identified using the iterative procedure presented in Section~\ref{sec.inc.RB}. All three proposed methods were considered and for each of these, three different sets of microstructures were considered: The first set of microstructures consisted of RVEs with only circular inclusions, the second set consisted of RVEs with only rectangular inclusions, and the third set was divided into equal parts, each part consisting of RVEs with either circular or rectangular inclusions\footnote{(i.e. each structures contained exclusively one of the two morphological prototypes and the same number of realizations for each prototype was enforced)}, respectively. Each type of microstructure was processed using each of the three incremental RB schemes introduced in Section~\ref{sec.inc.RB}. Hence, a total of nine different trainings were conducted, each using different randomly generated snapshots.

\begin{table}[!h]
	\caption{Data of the unsupervised learning (incremental RB identification) for the nine considered scenarios; the parameters $\varepsilon =0.025\,, n_{\rm c}=100 \text{ and }n_{\rm a}=75$ were used. Some numbers are rounded for easier readability.}\label{tab.method_comparison}
	\centering
	{\small 
\begin{tabular}{|c||c | c | c | c | c |c| }
\hline
 \begin{minipage}{1.2cm}\centering  Method \end{minipage}
&\begin{minipage}{1.4cm}\centering  final basis size \end{minipage}
&\begin{minipage}{1.7cm}\centering  snapshots \\ with $\mathcal P_{\delta}\geq \varepsilon$ \end{minipage}
&\begin{minipage}{1.7cm}\centering  snapshots \\ with $\mathcal P_{\delta}\leq \varepsilon$\end{minipage}
&\begin{minipage}{1.45cm}\centering  enrichment \\steps $n_{\rm it}$\end{minipage}
&\begin{minipage}{1.5cm}\centering  time $[$s$]$ \end{minipage}
&\begin{minipage}{3.0cm}\centering  used microstructures \end{minipage}\\[3mm]
\hline 
	A &	143&	150&	730&	4 & 20 &\\[2mm]
	B & 	80&	400&	2400&	7 & 70    	& \begin{minipage}{3cm}\centering \vspace{-6mm}\includegraphics[width=14mm,height=14mm]{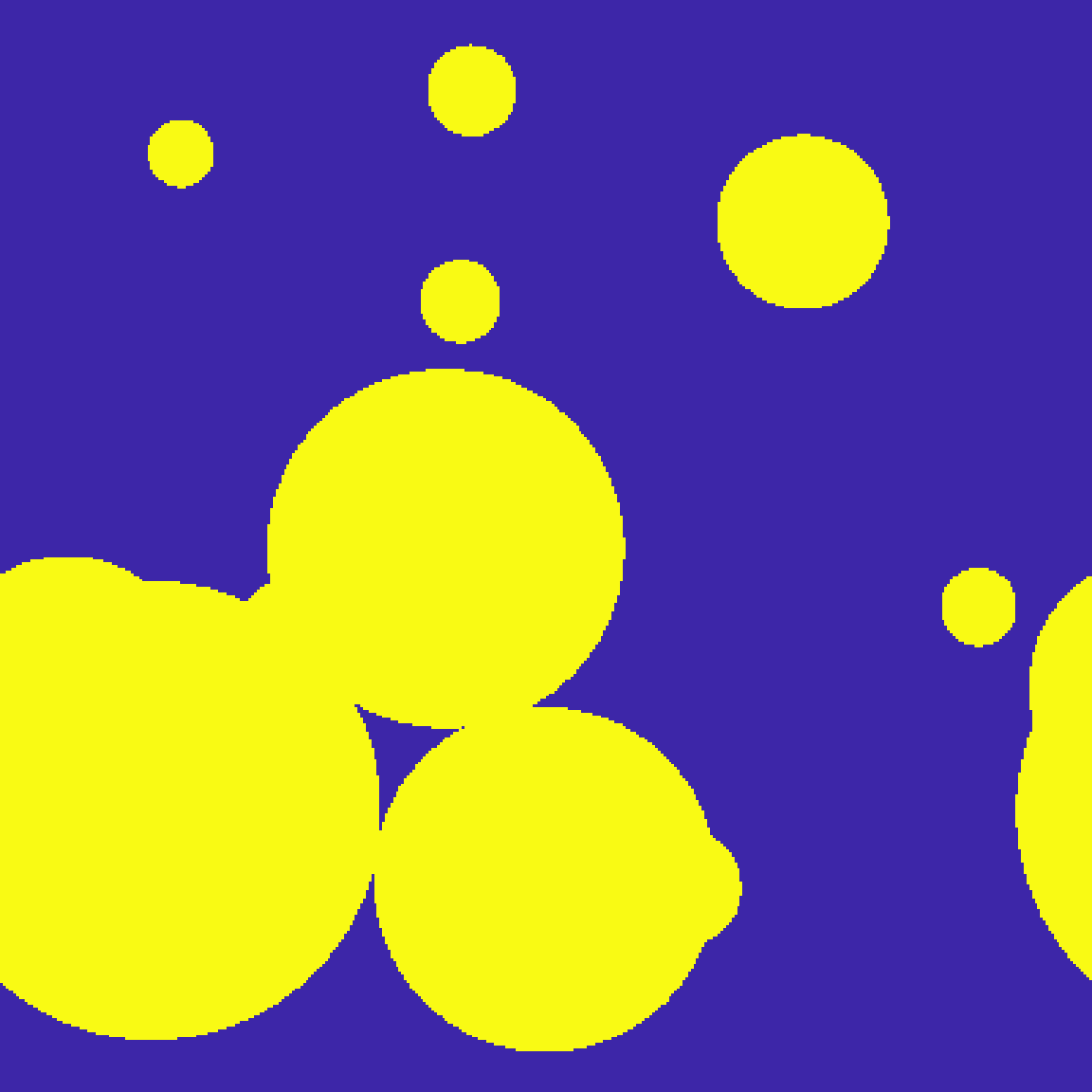}\vspace{-6mm}\end{minipage}\\[2mm]
	C &	96&	800&	7700&	12 & 200 &\\[2mm]
\hline
\hline
	A &	596&	670&	4500&	11 & 150 &\\[2mm]
 B & 	294&	2400&	12700&	34 & 500 	& \begin{minipage}{3cm}\centering \vspace{-6mm}\includegraphics[width=14mm,height=14mm]{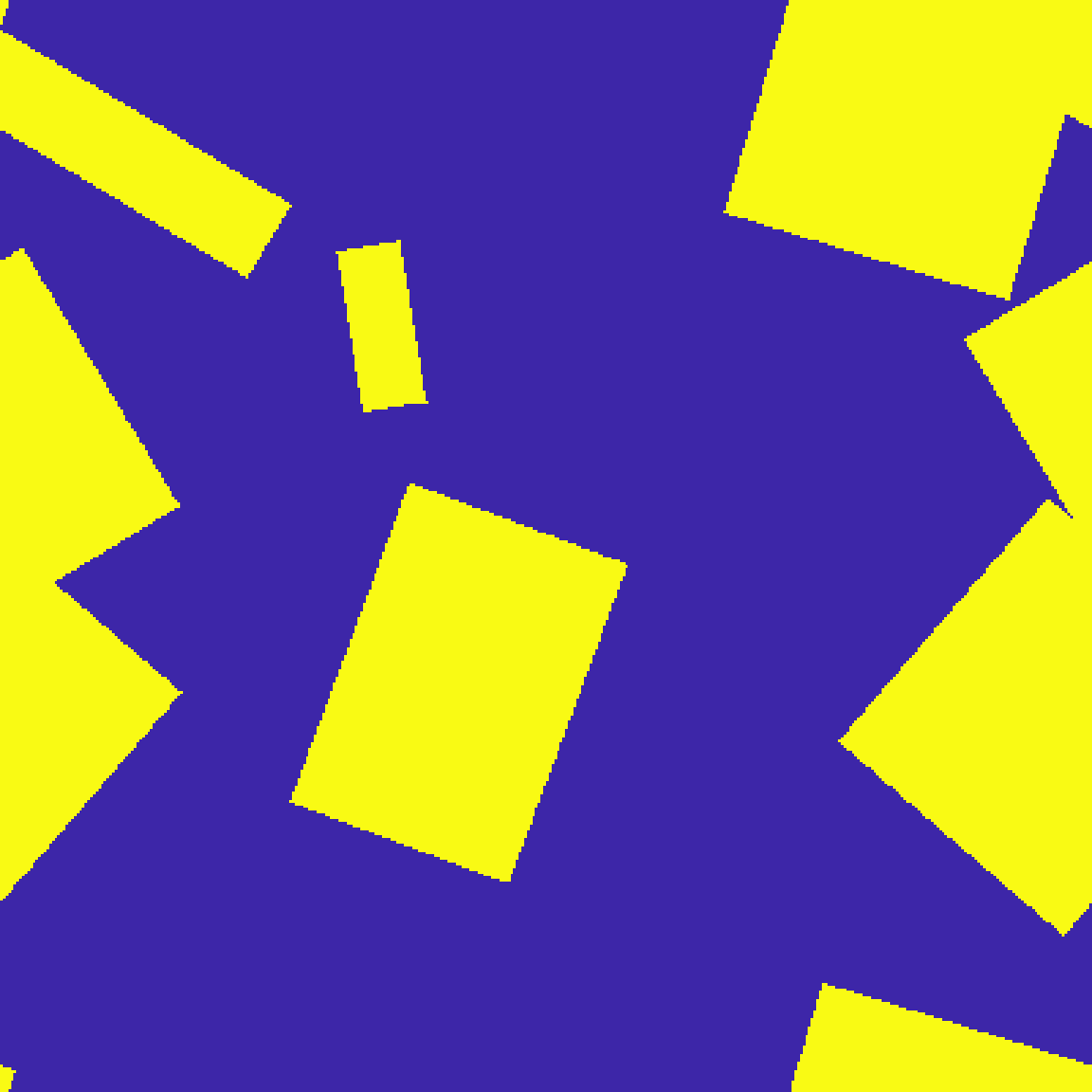}\vspace{-6mm}\end{minipage}\\[2mm]
	 C &	312&	2600&	16500&	37 & 550 &\\[2mm]
\hline
\hline
	A &	464&	560&	2900&	9 & 150 &\\[2mm]
 B & 	274&	2000&	16100&	29 & 500 	& \begin{minipage}{3cm}\centering \vspace{-6mm}\includegraphics[width=30mm,height=14mm]{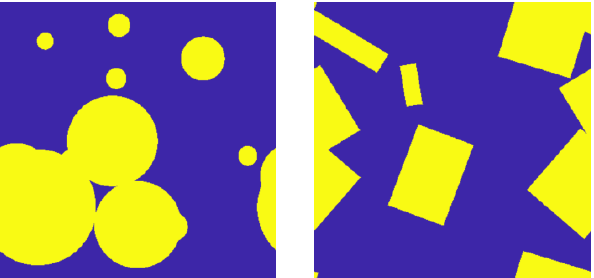}\vspace{-6mm}\end{minipage}\\[2mm]
	 C &	244&	1540&	8000&	22 & 280 &\\[2mm]
\hline
\end{tabular}
}
\end{table}

For the iterative enrichment process, the initial RB was computed from 200 snapshots $\ull S_{ 0}$. Thereafter, snapshots were randomly generated and processed by the enrichment algorithm sketched in \Figref{fig.it_b}. The number of snapshots per enrichment step has been set to $n_{\rm a}=75$ and the number of consecutive snapshots with $\mathcal P_{\delta}< \varepsilon$\,, used to indicate convergence, has been set to $n_{\rm c}=100$. The relative projection tolerance $\varepsilon=0.025$ was chosen. Note that this corresponds to the maximum value of the mean relative $\Vert\cdot\Vert_{L^2}$-error that is considered exact for the shifted snapshots. The actual accuracy in the reproduction of the 2PCF $c_2(\fr;b,b)$ is significantly lower than this (results are given in \Figref{fig.true_pd} below).

Key attributes for each of the nine trainings are provided in \Tabref{tab.method_comparison}. There is an obvious discrepancy between Method A and the remaining methods in basically all outputs. While Method A claims the lowest computing times, it yields approximately twice the number of modes. However, the number of snapshots needed is substantially lower which can be relevant if the generation of the synthetic microstructures is computationally involved. 

Note that methods B and C yield similar results, although for the rectangular and circular training method C needed significantly more snapshots, method B needed significantly more snapshots for the mixed training. The outliers in the number of snapshots needed are due to the randomness of the materials and the chosen convergence criterion. The resulting basis size of methods B and C indicate very similar results from these methods.

Also note that the calculation of the relative projection error $\mathcal P_{\delta}$ grows linearly with the dimension of the RB, i.e. the faster offline time of method A can quickly be compensated by the costly online procedure induced by the high dimension of the RB in comparison to the competing techniques.

\begin{figure}[b]
    \centering
	\includegraphics[width=1.0\textwidth]{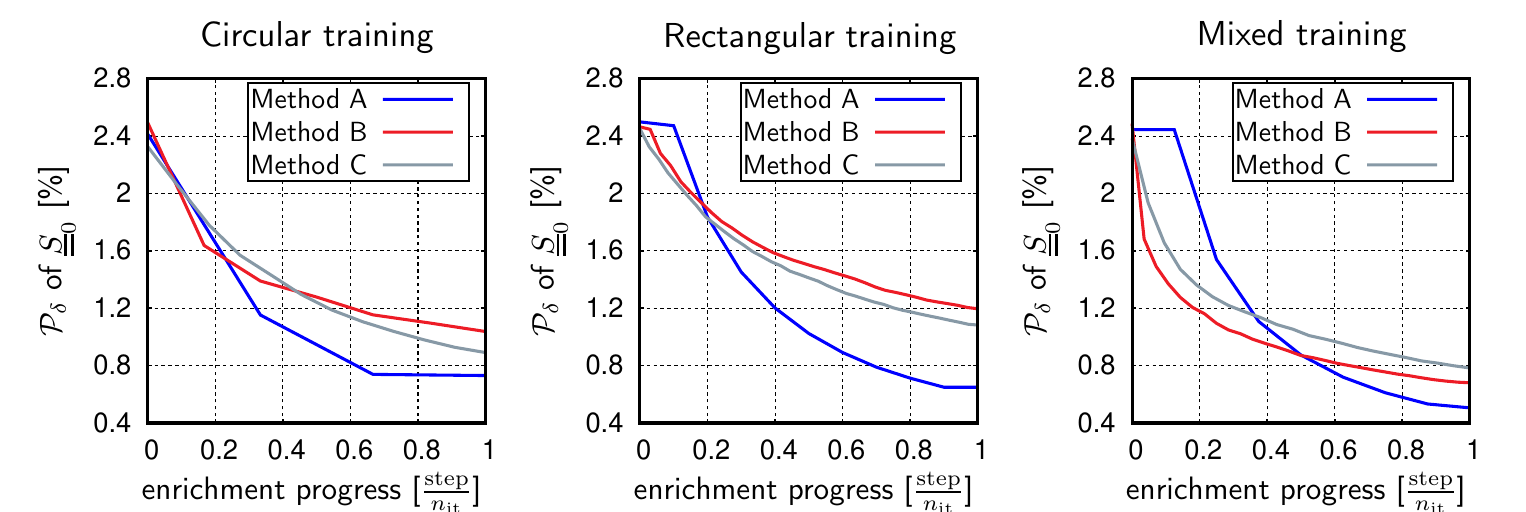}
	\caption{Development of the relative projection error $\mathcal P_{\delta}$ of the snapshots $\ull S_{0}$ with respect to the relative enrichment progress.}\label{fig.s_devel}
\end{figure}

\begin{figure}[t]
	\includegraphics[width=1.0\textwidth]{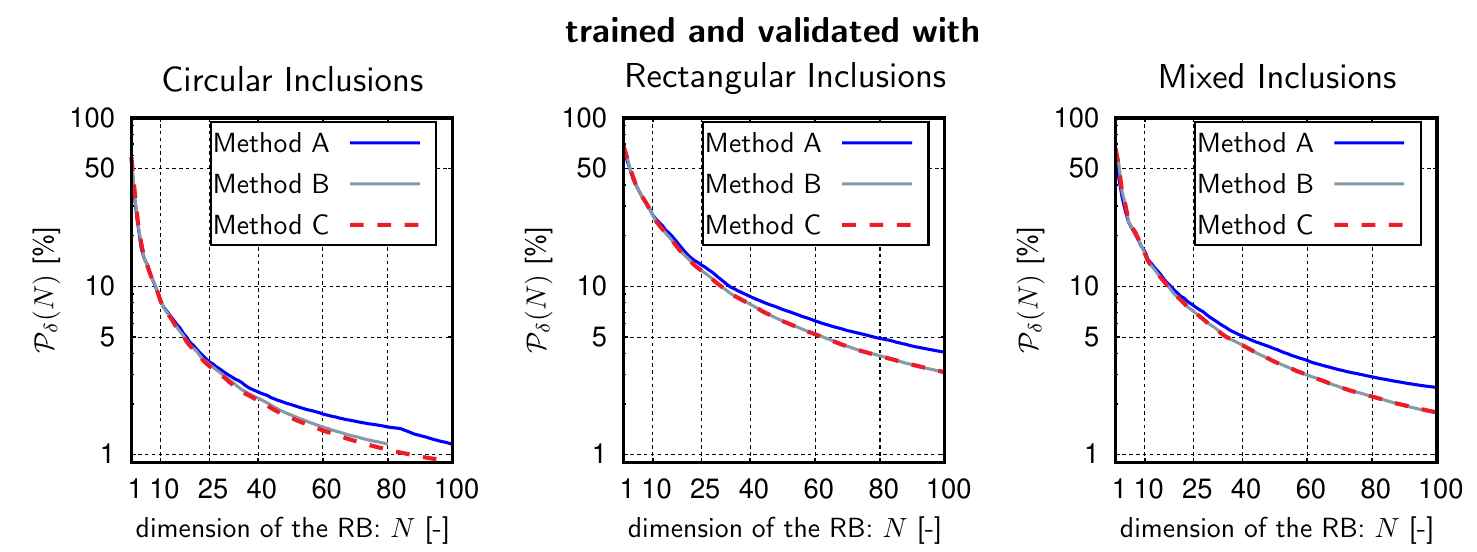}
	\caption{Relative projection error for three different microstructure classes as a function of the number of eigenmodes. The relative projection error is determined for a validation set of 1500 newly generated microstructures for each class.}\label{fig.basis_acc}
\end{figure}

To compare the accuracy of the resulting basis as well as during the training, the relative projection error $\mathcal P_{\delta}$ of the snapshots used for the original basis construction $\ull S_0$ are  plotted in \Figref{fig.s_devel}.

Since each training conducted a different amount of incremental updates, the abscissa has been rescaled such that the relative progress of the basis generation is shown vs. the number of enrichments divided by~$n_{\mathrm{it}}$ (\Figref{fig.s_devel} and \Tabref{tab.method_comparison}).

Method B and C yield again very similar results whereas method A achieves a lower projection error on convergence, but at the expense of a considerably improved dimension of the RB.
Since there seems to be an obvious correlation between resulting accuracy and the final basis size for the initial snapshots $\ull S_0$ (see \Figref{fig.s_devel}, \Tabref{tab.method_comparison}), the general quality for arbitrary stochastic inputs must be investigated. In order to quantify the quality of the RB, the accuracy can be expressed in terms of the relative projection error of approximating additional, newly generated snapshot data~$\ull{S}$ as a function of the method (A, B, C) and the number of modes~$N \geq 1$ via
\begin{equation}
	\mathcal P_{\delta}(N)= \sqrt{\frac {||\ull S-\ull B(:,1:N)\,\Tull B(:,1:N)\,\ull S||_F^2}{||\ull S||_F^2}} \label{eq.RB.acc}
\end{equation}
in Matlab notation.
\begin{figure}[!t]
\centering
	\includegraphics[scale=0.995,keepaspectratio=true]{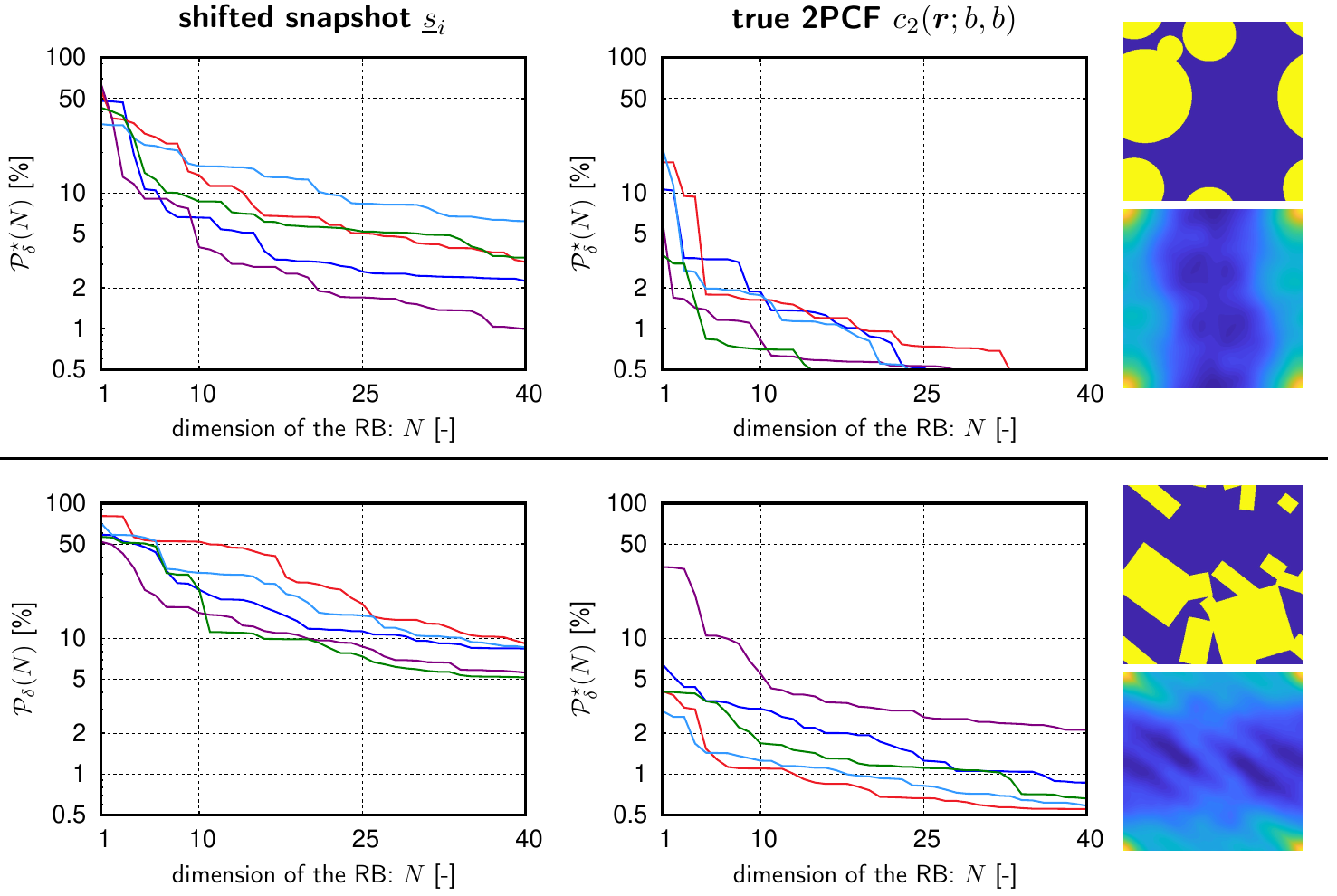}
	\caption{Using the RB of method C, the relative projection error on the shifted snapshot is given on the left for five random samples. For comparison the relative projection error of the reconstruction of the actual 2-point correlation function is given on the right for the same samples.}\label{fig.true_pd}
\end{figure}

\begin{figure}[!b]
\centering
	\includegraphics[width=1.0\textwidth]{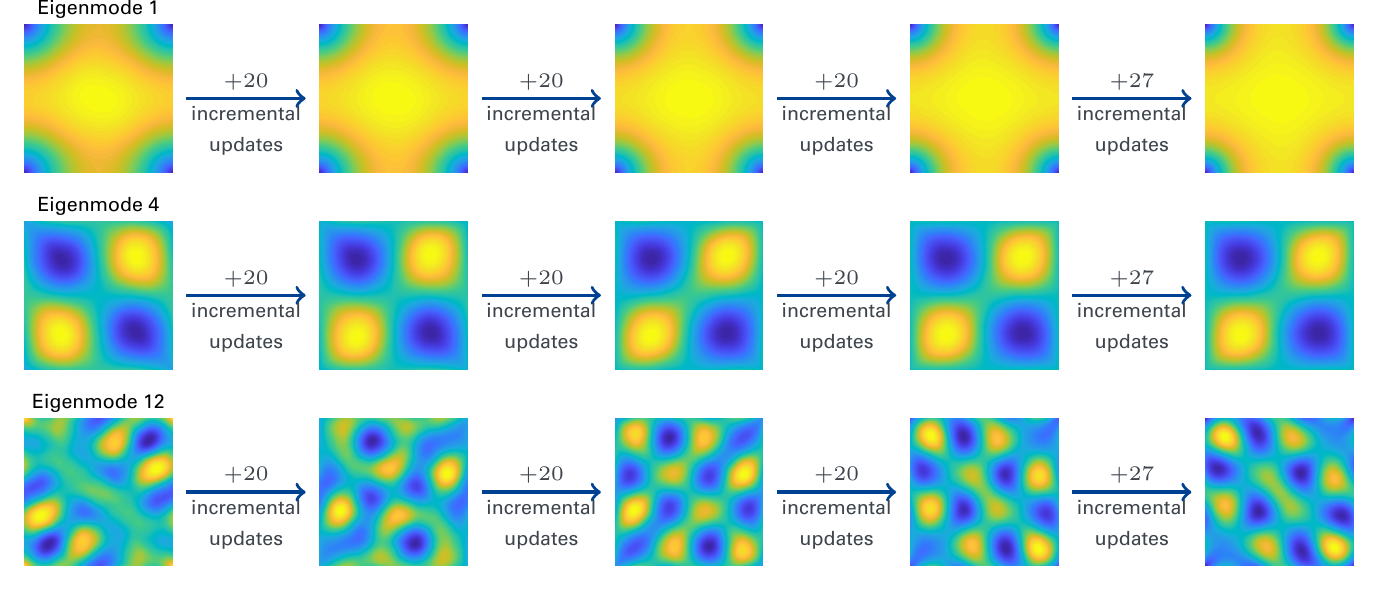}
	\caption{The development of a few selected eigenmodes over the enrichment are shown. Note that these results are generated with $n_{\rm a}=15$ and $\varepsilon =0.01$ using method C. The procedure comprised a total of 87 basis enrichments/adjustment.}\label{fig.eigenmodes}
\end{figure}
This measure captures to what extend the first $N$~basis functions represent the 2PCF of the underlying microstructure class. In the current work sets of 1500 newly generated snapshots assure an unbiased validation, i.e. the data was used in neither of the three training procedures. The results are stated in \Figref{fig.basis_acc}. Again, methods B and C yield similar results, achieving lower projection errors with fewer eigenmodes compared to Method A, i.e. the basis produced by method A cannot catch up with its two competitors. On a side note, the rectangular inclusions apparently lead to significantly richer microstructure information which can be seen by direct comparison of the left to the middle plot in \Figref{fig.basis_acc}. For methods B and C and for circular inclusions the relative error of 5\% is reached for approximately 15 modes while rectangular inclusions require more than 60 modes to attain a similar accuracy. This is supported also by the rightmost plot determined from a sort of blend of the two microstructural types.

 Since all of the previous error measures are given on the shiftedsnapshot according to \equref{eq.s_scaling}, the true relative projection error on the unshifted snapshot is also investigated as a function of the basis size. It describes the relative accuracy of the approximation of the 2PCF $c_2(\fr;b,b)$ as a function of the basis size. The errors in the shifted data (\Figref{fig.true_pd}, left) and the corresponding reconstructed 2PCF (\Figref{fig.true_pd}, right) for five individual randomly selected snapshots show that the actual relative error in the 2PCF reconstruction is below 5\% for 10 reduced coefficients even for the challenging rectangular inclusion morphology, while the error in the shifted and shifted snapshots is on the order of 50\%. This highlights the statement made earlier regarding the choice of $\varepsilon$ which is not directly the accepted mean error in the 2PCF, but only after application of the shift.  The high discrepancy in the two relative projection errors is due to the fact that the shifted snapshots fluctuate closely around 0, i.e. the homogeneous part of the 2PCF is obviously of high relevance.

The development, i.e. the stabilization of the mode shapes over the enrichment steps, of a few selected eigenmodes is shown in \Figref{fig.eigenmodes} using RVEs with circular inclusions for training of method~C. Similar results are expected for method~B, whereas for method~A the eigenmodes would remain unconditionally unchanged over the enrichment steps, i.e. a pure enlargement of the basis takes place. The faster stabilization of the leading eigenmodes indicates a quick stabilization of the lower order statistics of the microstructure ensemble, while the tracking of higher order fluctuations is more involved.

\subsection{Supervised Learning}
After the training of the RB, the input for the neural network, the feature vector $\ul{\xi}$ was derived using the 1- and 2-point spatial correlation functions of the $i$th RVE as
\begin{equation}
	\ul{\xi}_i=
		\begin{bmatrix}
		f_{b,i}\\		\Tull B\,\ul s_i \\		
		\end{bmatrix} \qquad \in \styy R^{(h+1)}\,.
\end{equation}
The size of the feature vector is determined by the amount of reduced coefficients~$1 \leq h \leq N$, i.e. the snapshot is projected onto the leading $h$~eigenmodes of~$\ull B$.

Since the inputs and outputs have a highly varying magnitude, they need to be shifted such that they are equally representative. Therefore, each entry of the feature vector is separately shifted and scaled such that its distribution of all samples has zero mean and a standard deviation of one. The output is shifted combinedly such that the mean of $\ul{\bar{\kappa}}_{\rm V}$ is $\ul{0}$. The transformed inputs and outputs are then given to the ANN for the training phase. Thus, the outputs of the ANN need to undergo an inverse scaling in order to yield the sought-after vector representation of the heat conduction tensor. These shifts and scalings need to be extracted from the available training data. Hence, every data set used for training purposes has its own parameters.

The training for the neural network has been conducted for all of the three microstructure classes, i.e. using only RVEs with circular inclusions, only RVEs with rectangular inclusions and lastly using RVEs with either circular or rectangular inclusions (the split was up to 60:40, which was randomly assigned before the training).  In order to derive the feature vector, the converged basis of method C has been used. Note that depending on the training set, only circular/rectangular or mixed inclusions contributed to the RB.

The training of the ANN was conducted with an \textit{early stop} algorithm: up to 10000 epochs were considered and the best--not necessarily last--parameter set has been saved. The decision on the \textit{best} ANN was taken on the basis of the cost function of the validation set (which came out of the same parameter range as the training set). Each data set consisted of 1500~samples. These were shuffled randomly and split into the training set ($n_t=1000$) and the validation set ($n_v=500$) before each ANN training.

\begin{figure}[!t]
	\includegraphics[width=1.0\textwidth]{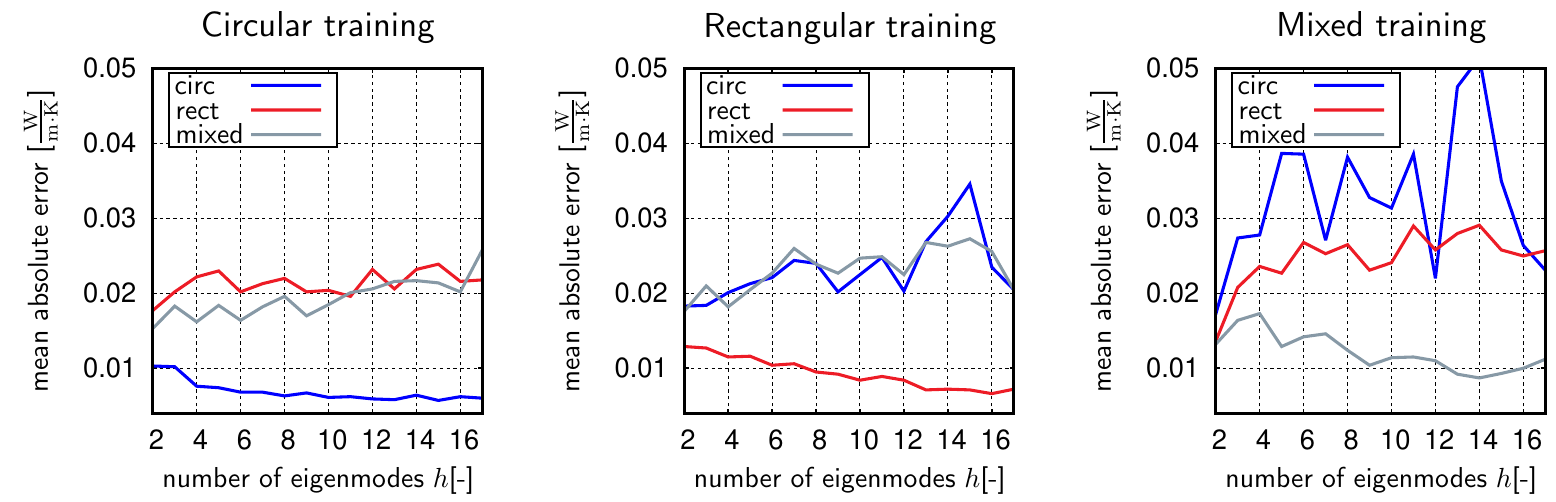}
	\vspace{4mm}
	\includegraphics[width=1.0\textwidth]{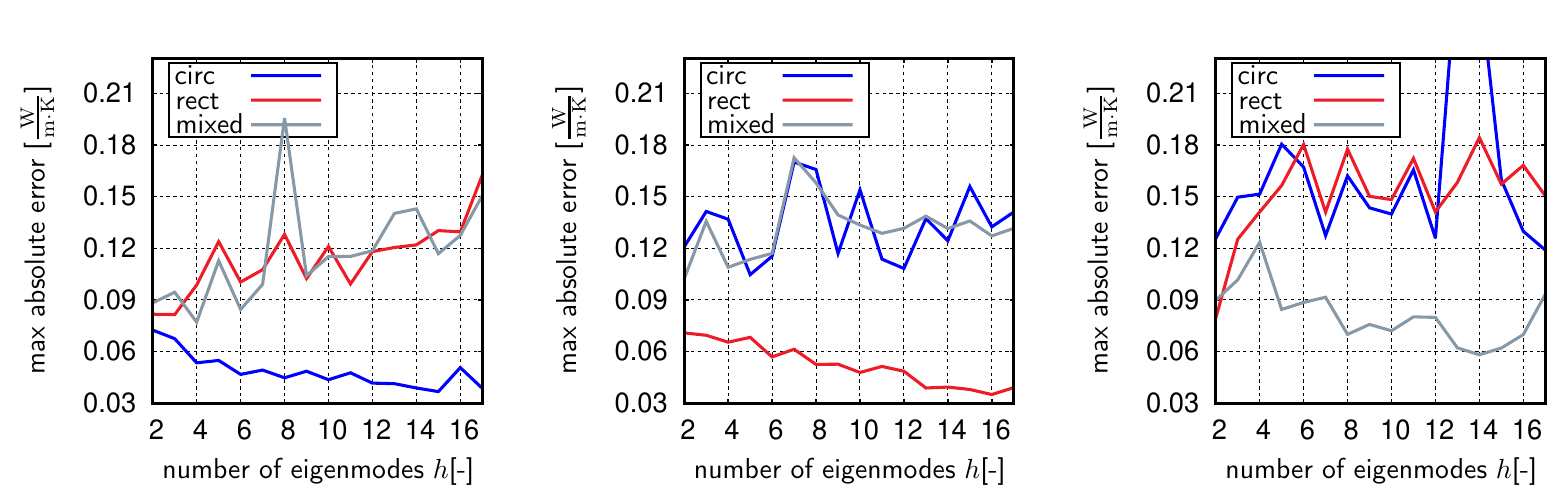}
	\caption{The given error measures over the test sets are shown for the ANN which achieved the lowest MSE (cost) on the training set for each number of reduced coefficients and training type.
    }\label{fig.red_coeff}
\end{figure}

In the following, the error measurements used and the term of unbiased testing refers to the prediction of 1500 unseen data points for each of the three microstructure classes named \textit{test sets}. In order to find a good  overall ANN, the network architecture has been intensely studied: the accuracy of the prediction after the training has been evaluated with various sizes of the feature vector, different network layouts and for different activation functions (\Figref{fig.red_coeff}).
The depicted error measure (\Figref{fig.red_coeff}) is the mean/max absolute derivation, i.e. the absolute error (corresponding to the Euclidean norm of the vector-valued error) for the prediction and the actual heat conductivity of the microstructure of $\ul{ \ol {\kappa}}_{\rm V}$ over each test set.

\begin{figure}[b!]
	\includegraphics[width=1.0\textwidth]{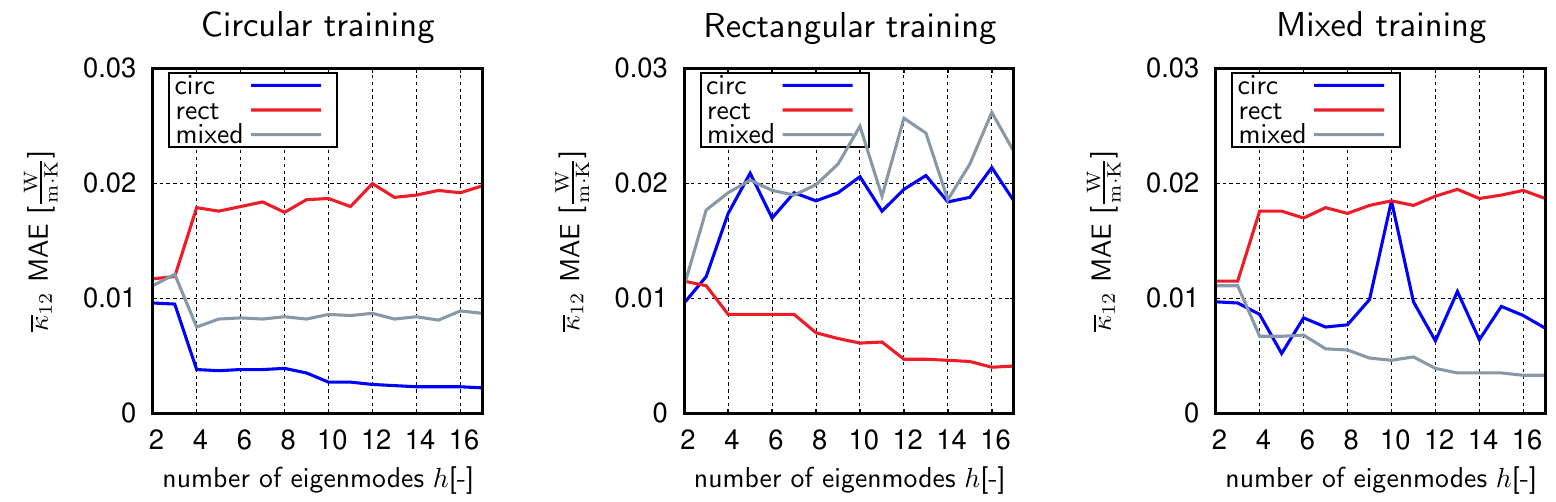}
	\caption{The mean absolute error (MAE) of $\ol{\kappa}_{12}$ is given for each of the trainings and test sets.
    }\label{fig.k12_coeff}
\end{figure}

The conductivity~$\ol{\kappa}_{12}$ fluctuates mildly around zero for all inputs. In order to accurately capture this fluctuation, only the specific training and RB dimensions of four or higher ($h\geq 4$) are required cf. \Figref{fig.k12_coeff}. However, the error of $\ol{\kappa}_{12}$ for other inclusion morphologies than that of the training increase with~$h$, albeit the values can be considered small in comparison to the $\ol{\kappa}_{11}$ and  $\ol{\kappa}_{22}$ errors.

Since the trend in \Figref{fig.red_coeff} of the rectangular and mixed training looks promising, a further study using more reduced coefficients has been conducted, however the results for retangular training showed a stronger tendency towards overfitting. Rectangular training with $17\leq h\leq 25$ had very similar results to circular training with $12\leq h\leq 17$ reduced coefficients. For the mixed training the overall results worsened with a higher amount of reduced coefficients and the ANN did not seem to find any mapping generalizing the property linkage for both types of deployed RVE with the given training data.

The increased overfitting with respect to the training microstructure class for increased dimension of the feature vector can be explained by the rather limited number of 1000 input samples. For instance, considering a six-dimensional feature vector induces that for the rather limited number of ten samples per independent direction a total of 10\textsuperscript{6} data points would be needed. The dilemma is that each of the computations is expensive, particularly when considering three-dimensional simulations, i.e. millions of samples can not be realized in practice. Therefore, the number of samples must be rather low which could be a limiting factor in view of the number of features that can be accounted for.

Overall, a correlating trend with the accuracy for the RB (\Figref{fig.basis_acc}) and the behaviour of the ANN training could be seen (\Figref{fig.red_coeff}). Investigating the circular training, with more specific information about the 2PCF available, a slightly better mapping is found for RVEs with circular inclusions, whereas the prediction for the other two microstructure classes slightly worsen. Similarly with the rectangular training, more reduced coefficients increase the fit on the RVE with rectangular inclusions, though more reduced coefficients are required to deliver the same accuracies as the circular training, as more number of eigenmodes (of the basis) are required to yield approximately the same relative projection error. 

\begin{figure}[b!]
	\includegraphics[width=1.0\textwidth]{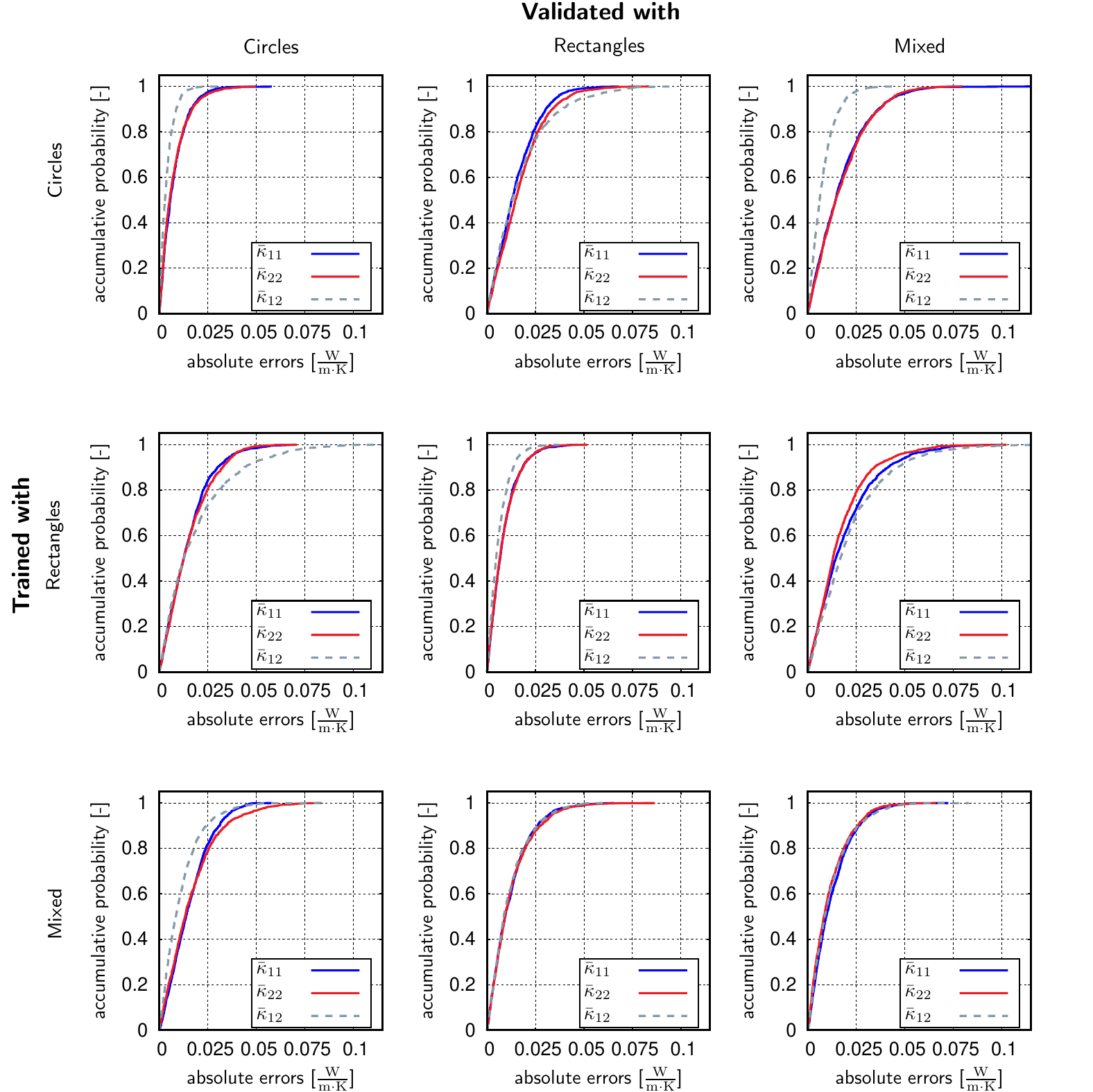}
	\caption{Results for the best of all tested ANN for the test sets. The graphs represent probability distribution of the absolute error in the components of $\ul{\bar{\kappa}}$. }\label{fig.nn_acc}
\end{figure}

When training with both types of RVEs (i.e. for the mixed input set), the training seemed more random than the others. Some resulting ANNs (which are not shown) had more of the property of a circular training, whereas some were more like the rectangular training. Although the ANN was trained with general RVEs, it overall failed to give good predictions for every type of RVE when using more than two or three reduced coefficients. After training with the mixed set, the prediction of $\bar{\kappa}_{11}$ and $\bar{\kappa}_{22}$ was generally better for RVEs with rectangular inclusions, whereas $\bar{\kappa}_{12}$ was better for RVEs with circular inclusions. During the training, the ANN found most likely some strange local minimum which fitted the training data quite well, however it was not a general mapping for all the microstructure classes, which hints at a too low number of input samples, as discussed earlier.

The ANN architecture did not seem to have a big impact on the quality of the prediction, there have been ANNs with a single hidden layer and six hidden neurons which delivered results comparable to an ANN comprising six hidden layers with more than 100 hidden neurons.

The used activation functions were the sigmoid, relu, tanh and softplus, where only some combinations delivered poor results. Not a clear trend of ANN architecture and quality of prediction could be seen and, consequently, the best ANN were randomly found based on the lowest error on the test set.\\
The prediction accuracies for each test set of three differently trained ANNs, which have been deemed the \textit{best}, is given in \Figref{fig.nn_acc}. As to be expected, the lowest errors are achieved on the diagonal, i.e. training set $\widehat=$ validation set.\\ 
The training and architecture of the best ANNs in \Figref{fig.nn_acc} had the following properties:\\
\begin{equation*}
	\begin{array}{l@{\hspace{-4mm}}ll}
		&\bullet\,\text{Circular training:} &\text{$h=6$; 4570 epochs; 2 hidden layers } \\
		&  & \text{\{7, 39\} hidden neurons; activation functions \{relu, softplus\}}\\[2mm]
		&\bullet\, \text{Rectangular training:}& \text{$h=10$; 560 epochs; 3 hidden layers } \\
		&  &  \text{\{13, 82, 25\} hidden neurons; activation functions \{sigm, tanh, sigm\}}\\[2mm]
		&\bullet\, \text{Mixed training:} &\text{$h=2$; 4760 epochs; 1 hidden layer } \\
		&  &  \text{\{20\} hidden neurons; activation function \{relu\}}\\
	\end{array}
\end{equation*}

For an easier readability, the percentage mean and max errors for each ANN training and prediction are given in \Tabref{tab.nn_percent}, again using the same three ANNs which have been explicitly shown. Note that since the values of $\bar{\kappa}_{12}$ vary closely around 0 (\Figref{fig.rangekappa}), relative errors are not sensible for the quantity of interest.

\begin{table}[h!]
	\caption{Percentage errors for $\bar{\kappa}_{11}$ and $\bar{\kappa}_{22}$ given for each of the best ANNs, validated on every test set.}\label{tab.nn_percent}
	\centering
	\small{

\begin{tabular}{|c||c | c  c | c  c | c c| }
	\cline{3-8}
		\multicolumn{2}{c|}{} 	& &\multicolumn{4}{c}{validated with} &  \\[2mm]
	\cline{3-8}
		\multicolumn{2}{c|}{} & \multicolumn{2}{c|}{circles} & \multicolumn{2}{c|}{rectangles} & \multicolumn{2}{c|}{mixed} \\
		\cline{3-8} 
		\multicolumn{2}{c|}{}&&&&&&\\[-3mm]
		\hline
 \begin{minipage}{1.2cm}\centering  trained with \end{minipage}
&\begin{minipage}{1.4cm}\centering  error\\ measures \end{minipage}
	&\begin{minipage}{1.7cm}\centering  $\ol{\kappa}_{11}$ \end{minipage}
	&\begin{minipage}{1.7cm}\centering  $\ol{\kappa}_{22}$\end{minipage}
	&\begin{minipage}{1.45cm}\centering  $\ol{\kappa}_{11}$ \end{minipage}
	&\begin{minipage}{1.5cm}\centering  $\ol{\kappa}_{22}$ \end{minipage}
	&\begin{minipage}{1.0cm}\centering  $\ol{\kappa}_{11}$\end{minipage}
	&\begin{minipage}{1.5cm}\centering  $\ol{\kappa}_{22}$ \end{minipage}\\[3mm]
\hline 
	\multirow{2}{*}{circles} &   mean [\%]&      1.7&	 1.7&         3.5&	   3.9&          3.9&	      3.8  \\[2mm]
				 &   max [\%]&     13.9&	12.2&        16.4&	  24.6&         27.1&	     16.4  \\[2mm]
				 \hline
	\multirow{2}{*}{rectangles} &mean [\%]&      3.4&	 3.6&         1.9&	   1.9&          4.2&	      3.7  \\[2mm]
				 &   max [\%]&     19.5&	19.9&         8.9&	  10.9&         24.2&	     21.0  \\[2mm]
				 \hline
	\multirow{2}{*}{mixed} &     mean [\%]&      3.4&	 3.6&         2.6&	   2.6&          2.7&	      2.4   \\[2mm]
				 &   max [\%]&     13.7&	19.9&        12.8&	  15.6&         15.6&	     14.7  \\[2mm]
\hline
\end{tabular}

	}
\end{table}

A GUI code is provided in \href{https://github.com/J-lissner/img---kappa}{Github}, where the user can choose between the three proposed ANN, the input for the prediction is a $400\times 400$ image in matrix format written in a text file or a TIFF image and the output is the prediction for the heat conduction tensor as described above. In order to compile the code, Python3 with Tensorflow is required. Some exemplary RVE with their respective heat conductivity are uploaded in a subfolder.

\section{Conclusion}
\subsection{Summary and Concluding Remarks}
The computational homogenization of highly heterogeneous microstructures is a challenging procedure with massive computational requirements.
In the present study a method to efficiently and accurately predict the heat conductivity for any RVE with the image and no further information is proposed. Key ideas of the Materials Knowledge System (MKS) \cite{fast2011formulation,kalidindi2012computationally} have been adopted in the sense that a subset of the POD compressed 2-point correlation function is used to identify a low-dimensional microstructure description. In contrast to \cite{fast2011formulation} the 2PCF is not truncated to a small neighborhood, but the full field information is considered. Similar to other works related to the MKS \cite{brough2017materials}, a truncated PCA of the 2-point information is used to extract microstructural key features.

However, the classical truncated PCA used, e.g., in \cite{brough2017materials} is not applicable to the considered rich class of microstructures due to the high number of needed samples and the related unmanageable computational resources. Therefor, our proposal is founded on a novel incremental procedure for the generation of the RB of the 2PCF. Similar techniques have not been considered in the literature to the best of the authors' knowledge. The shifting of the images of 2PCF before entering the POD is another feature that can help in reducing the impact of the inclusion volume fraction, i.e. the shifted function has zero mean.

Other than in \cite{fast2011formulation} no higher-order statistics are used. This is by purpose as the selection of the relevant entries of the higher order PCF is ambiguous and a challenge in itself. Instead, the present study focus on the variability of the input images in terms phase volume fractions in a broad range (20-80\%) alongside topological variations (impenetrable, partial overlap, unrestricted placement) and different morphologies (circles and rectangles). Generally speaking, a much higher microstructural variation is accounted for, than in previous studies. Therefore, the current study also investigates how the proposed technique and similar MKS related approach can possibly generalize towards truly arbitrary input images (e.g. millions of snapshots) in order to built a \textit{super-database}.

In order to cope with the variability of the 2PCF, the truncated PCA or snapshot POD effected during unsupervised learning phase is replaced by novel incremental procedures for the construction of a small-sized reduced microstructure parameterization. Three increment POD methods are proposed and their results are compared regarding the computational effort, the projection accuracy of the snapshots and the quality of the basis in view of capturing random inputs.

The learned reduced bases are used to extract a low-dimensional feature vector which denotes the input of a fully connected feed forward artificial neural network. The ANN is used to predict the homogenized heat conductivity of the material defined by the microstructure. The mean relative error of the surrogate is lower than 2\% for the majority of the considered test data. This is remarkable in view of the phase contrast~$R=5$ and the particle volume fractions ranging from 0.2-0.8, as well as morphological and topological variations. Further, an immense speedup in computing time is achieved by the surrogate over FE or FFT simulations.

Importantly, the presented methodology can immediately be adopted to different physical settings such as thermo-elastic properties, fluid permeability, dielectricity constants etc. The same holds for three-dimensional problems. However, the limited number of samples in 3d could be problematic as more features are likely required to attain a sufficiently accurate RB.

\subsection{Discussion and Outlook}
A weakness of the current approach remains the computational complexity of the method:
Although the feature vector is rather low-dimensional, it requires the evaluation of the 2PCF using the FFT which is of complexity $\cO(n \, \text{log}(n) )$ where $n$ is the number of pixels/voxels in the image.
In order to extract the reduced coefficient vector from the 2PCF, the latter must be projected onto the RB. This operation scales with $\cO( n \, N )$. Both operations are at least linear to the number of pixels/voxels of the image which can be critical, especially in three-dimensional settings. Therefor, future investigations aiming at a reduced effort for gathering the features are required in our opinion.

Another extension of the current scheme could account for variable phase contrast~$R$ which was fixed as $R=5$ in this work. Thereby, the dimension of the feature vector will be incremented which can add to the already existing data scarcity dilemma observed when considering a sufficient number of reduced coefficients: the number of input samples for the supervised learning should grow exponentially with the dimension of the feature vector. In the authors opinion this dependence is the most pronounced short-coming of the method and future studies should focus on limiting the number of required input samples in order to fight the curse of dimensionality: more reduced coefficients require an exponential growth in the available data, making the offline procedure unaffordable, today.


Advantages of the current scheme comprise the independence of the underlying simulation scheme. This does allow for heterogeneous simulation environments, the use of commercial software, multi-fidelity input data and blended sources of information (e.g. in silico data supported by experimental results).

\vspace{6pt} 



\authorcontributions{
   Conceptualization, J.L. and F.F.; Data curation, J.L.; Formal analysis, F.F.; Funding acquisition, F.F.; Investigation, J.L. and F.F.; Methodology, J.L. and F.F.; Project administration, F.F.; Resources, F.F.; Software, J.L.; Supervision, F.F.; Validation, J.L.; Visualization, J.L.; Writing -- original draft, J.L.; Writing -- review \& editing, J.L..
}

\funding{This research was funded by Deutsche Forschungsgemeinschaft (DFG) within the Emmy-Noether programm under grant DFG-FR2702/6 (contributions of F.F.).}

\acknowledgments{Support from Mauricio Fern\'andez on the implementation and layout for the training of the Artificial Neural Networks using Google's TensorFlow is highly appreciated. Stimulating discussions within the Cluster of Excellence SimTech (DFG EXC2075) on machine learning and reduced basis methods are highly acknowledged.}

\conflictsofinterest{The authors declare no conflict of interest.}
\reftitle{References}
\end{document}